\documentclass[12pt]{iopart}

\usepackage{amssymb}
\usepackage{color}
\usepackage{graphics, graphicx}
\usepackage{bbold}
\usepackage{psfrag}
\usepackage{mathcomp}
\usepackage{subfigure}
\usepackage{verbatim}
\usepackage{float}
\usepackage{iopams}
\begin{document}


\title[Driven-dissipative many-body pairing states]{Driven-dissipative many-body pairing states for cold fermionic atoms in an optical lattice}

\author{W Yi$^{1,2,3}$, S Diehl$^{1,2}$, A J Daley$^{1,2,4}$ and P Zoller$^{1,2}$}
\address{$^1$ Institute for Quantum Optics and Quantum Information of the Austrian Academy
of Sciences, A-6020 Innsbruck, Austria}
\address{$^2$ Institute for Theoretical Physics, University of Innsbruck, A-6020 Innsbruck, Austria}
\address{$^3$ Key Laboratory of Quantum Information, University of Science and Technology of China,
CAS, Hefei, Anhui, 230026, People's Republic of China}
\address{$^4$ Department of Physics and Astronomy, University of Pittsburgh, Pittsburgh, Pennsylvania 15260, USA}
\ead{wyiz@ustc.edu.cn}

\begin{abstract}
We discuss the preparation of many-body states of cold fermionic atoms in an optical lattice via controlled dissipative processes induced by coupling the system to a reservoir. Based on a mechanism combining Pauli blocking and phase locking between adjacent sites, we construct complete sets of jump operators describing coupling to a reservoir that leads to dissipative preparation of pairing states for fermions with various symmetries in the absence of direct inter-particle interactions. We discuss the uniqueness of these states, and demonstrate it with small-scale numerical simulations. In the late time dissipative dynamics, we identify a ``dissipative gap'' that persists in the thermodynamic limit. This gap implies exponential convergence of all many-body observables to their steady state values. We then investigate how these pairing states can be used as a starting point for the preparation of the ground state of Fermi-Hubbard Hamiltonian via an adiabatic state preparation process also involving the parent Hamiltonian of the pairing state. We also provide a proof-of-principle example for implementing these dissipative processes and the parent Hamiltonians of the pairing states, based on $^{171}$Yb atoms in optical lattice potentials.
\end{abstract}
\maketitle

\section{Introduction}

Quantum simulation using cold atoms in an optical lattice typically requires cooling to low temperatures to see interesting quantum phases with strong correlations \cite{ketterlecooling,jasonhopnas,trebst,LukinAF,Greinerqs}. An important example in this context is the quantum simulation of the 2-D Fermi-Hubbard Model (FHM) using cold fermionic atoms in an optical lattice \cite{Lukin02}. As a crucial first step toward this goal, various experimental groups have been successful in realising the 3-D Fermi-Hubbard model in such a system \cite{Ketterle06,Scheider,Jordens,Jordens2}. But cooling of the system below the critical temperature, and thus into the phases of interest, turns out to be difficult with the conventional cooling schemes ($T/T_F\sim 0.25$ \cite{Jordens,Jordens2}, where $T_F$ is the Fermi temperature). On the other hand, the atomic physics of these systems opens the possibility for a different approach to the production of interesting many-body states, specifically to dissipatively drive the system into steady states with the desired coherence and symmetry properties by careful engineering of a reservoir \cite{prosen1,Diehl08,Kraus,Verstraete09,prosen2,prosen3,tomadin,Diehl10,dwaveprl}. In this paper, we will discuss schemes by which fermions in a 2-D lattice potential can be dissipatively driven into pairing states with non-trivial correlations even in the absence of attractive interactions, extending our previous work \cite{dwaveprl}. These many-body pairing states can then also be used as a low-entropy starting point for efficient adiabatic passages, through which the true many-body ground state of many-body models such as the Fermi-Hubbard model might be reached \cite{trebst,LukinAF,Greinerqs}.

For Markovian dissipative processes, the coupling with the reservoir can be modeled by a set of dissipative quantum jump operators, which, when chosen appropriately, drive the system towards a pure steady state with zero entropy starting from \emph{any} given initial state. Similar ideas have been applied recently for the dissipative preparation of many-body states in optical lattices, e.g. a Bose-Einstein condensate for bosons \cite{Diehl08}, an $\eta$-condensate for fermions \cite{Kraus}, and most recently, d-wave pairing states in optical lattices \cite{dwaveprl} as well as topological phases \cite{Diehl11}. Here, we will focus on the dissipative generation of pairing states for fermions in an optical lattice. Starting from the general principles, we will show in detail how relative phases between the atoms can be imposed by engineering the jump operators, which in turn allows us to prepare many-body pairing states with specific spatial symmetries, e.g. p-wave and d-wave pairing symmetries. Note that in contrast to the equilibrium states of typical non-dissipative processes, the dissipative dynamics is described by a master equation, and the final state that we aim to prepare appears as the steady state of the dynamical process, providing a targeted many-body cooling protocol. As a conceptually important result, we observe that the imaginary spectrum of the effective Hamiltonian of the master equation has a gap which persists in the thermodynamic limit and we refer to it as a ``dissipative gap'', due to the formal analogy to the pairing gap of conventional BCS paired states \cite{leggettbook}. Physically, it leads to exponential convergence of many-body observables to their steady state values in the late time dissipative dynamics. This is a unique feature of the dissipative preparation of paired states with fermions.

We then propose an adiabatic process to connect the steady state of the dissipative process with the ground state of a Hamiltonian with similar symmetry properties. In the case of ideal adiabaticity, the system remains in a pure state and evolves into the ground state of the Hamiltonian without going through any Landau-Zener crossings. We consider this in the context of the Fermi-Hubbard model, where a dissipative process can be used to create \emph{dissipatively bound fermion pairs} which have d-wave symmetry. As we will see later, this BCS-type mean-field d-wave state lacks the strong correlations associated with the repulsive Fermi-Hubbard model, as the double-occupancy is not projected out by the dissipative dynamics. We will show that strong correlations can be built up through proper adiabatic passage in the sense that the steady state can be connected with the ground state of the Fermi-Hubbard model efficiently. The central question in the study of the 2-D FHM is whether the ground state exhibits d-wave superconductivity/superfluidity away from half-filling, and thus captures the universal properties shared by the high-Tc superconducting materials \cite{Anderson87,FCZhang,CGros,readgreen,Altman,Gurarie,Tewari,Haas,Trivedi}. If this is indeed the case, the coherence and the d-wave symmetry should be conserved during the adiabatic process. 

Finally, we note that a key physical ingredient for the jump operators for pairing states of fermions is the Pauli exclusion principle \cite{sandner2011}. Based on this understanding, we propose to implement the jump operators stroboscopically using alkaline-earth-like atoms. The existence of long-lived meta-stable states and rich level structures in these atoms provides us with the freedom to engineer various dissipative processes.

The paper is organised as follows: In Sec. II, we discuss the general formalism for jump operator engineering, and derive both the fixed-number and the fixed-phase jump operators for pairing states with different symmetries; in Sec. III, we study the uniqueness of the steady state under these jump operators for various cases both from the symmetry perspective and with small scale numerical calculations; we then derive a mean-field theory in Sec. IV for the master equation describing the dissipative dynamics, where a ``dissipative gap''  for the pairing states -- a minimal damping rate for the many-body relaxation -- is shown to emerge; in Sec. V, having established the dissipative preparation of pure steady states, we illustrate how we may start from these initial states to prepare the ground state of Hamiltonians with similar symmetries, for example the FHM, via an efficient adiabatic process; we discuss the implementation of the jump operators using alkaline-earth-like atoms in Sec. VI; finally we summarise and discuss other possible applications of our dissipative state preparation setting.

\section{General principles of reservoir engineering}\label{secgeneral}

Closed systems, where the energy and particle number are conserved, are modeled by a Hamiltonian, which determines the ground state and the dynamics via the Schr\"odinger equation. In this context, engineering states by realisation of a particular Hamiltonian, for which the desired state is the ground state is often discussed. It is then natural to ask similar questions for open quantum systems, where dissipative processes interrupt coherent evolution. In particular, one can consider engineering the dissipative processes so that their back-actions project the system into the desired subspace of the complete Hilbert space. For Markovian dissipative processes, as appropriate for the systems considered below, the dynamics of the density matrix for an open system is described by a master equation,
\begin{equation}
\frac{\partial \rho}{\partial t}=\mathcal{L}\rho\equiv -iH_{\mathrm{eff}}\rho+i\rho H_{\mathrm{eff}}^{\dagger}+\kappa\sum_{\ell}j_{\ell}\rho j_{\ell}^{\dagger},
\end{equation}
where the non-Hermitian effective Hamiltonian is given by
\begin{equation}
H_{\mathrm{eff}}=H-\frac{\mathrm i}{2}\kappa\sum_{\ell}j_{\ell}^{\dagger}j_{\ell}.
\end{equation}
Here, $\{j_{\ell}\}$ are non-Hermitian Lindblad operators reflecting the system-bath coupling with rate $\kappa$ \cite{Lindblad}. The Hamiltonian $H$ generates unitary evolution, and describes non-dissipative processes of the system. Although $H$ does not have to vanish in general, we will assume $H=0$ throughout the state preparation discussion, i.e. the final state is prepared via purely dissipative processes. For fermions loaded into an optical lattice, this can be achieved by increasing the lattice depth to freeze out the kinetic motion, while tuning the inter-particle interaction to zero, e.g. via a Feshbach resonance. Note that by considering a purely dissipative process, we can avoid competition between the Hamiltonian and the dissipative dynamics, which is present when the dark state is not an exact eignestate of the Hamiltonian. In particular, the pairing states that we aim to prepare are not exact eigenstates of the FHM in general.

\begin{figure}[tb]
\centerline{\includegraphics[width=10cm]{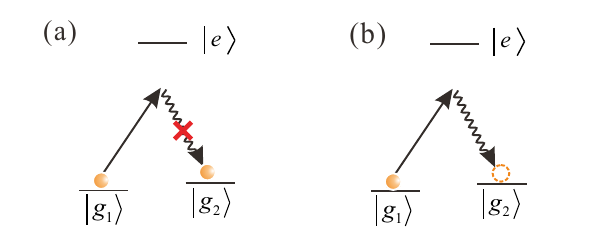}}
\caption{Illustration of the Pauli blocking in the optical pumping process of a three level system. (a) The spontaneous decay from the excited state $|e\rangle$ to the ground state $|g_2\rangle$ is blocked due to Pauli exclusion, leaving the system unchanged; (b) The decay channel is not blocked, and the population in $|g_1\rangle$ is transferred to the state $|g_2\rangle$. }
\label{Fig1_Jump_Operators}
\end{figure}

In the quantum trajectory picture, the system wavefunction $|\psi(t)\rangle$ of a given trajectory evolves according to the non-Hermitian Hamiltonian $|\psi(t)\rangle\propto e^{-iH_{\mathrm{eff}}t}|\psi(0)\rangle$, and is punctuated by the quantum jump $|\psi(t)\rangle\rightarrow j_{\ell}|\psi(t)\rangle$ with rate $\kappa\left\Vert j_{\ell}|\psi(t)\rangle\right\Vert^{2}$. The time-dependent density matrix is then determined by $\rho(t)=\langle|\psi(t)\rangle\langle\psi(t)|\rangle_{\mathrm{stoch}}$
\cite{zollerbook}, where the average runs over all trajectories.  In this picture, we see that for any state satisfying $j_{\ell}|\mathrm{BCS}_{N}\rangle=0$ $\forall\ell$, the quantum jumps will never project it to other states. These states are therefore ``dark states" of the jump operators, and are necessarily steady states of the master equation evolution \cite{Diehl08,Verstraete09}. If no other stationary solutions exist, the system will be driven to this state by the dissipative dynamics regardless of its initial conditions [11]. Similar ideas have been exploited for the dynamical preparation of a BEC for cold bosons loaded into an optical lattice, where the jump operators give rise to quasi-local phase-locking mechanism, which eventually leads to the condensation of the bosons in the lattice [10]. Here, we will focus on the preparation of pairing states for fermions in an optical lattice. For the fermionic case here, in addition to the phase-locking mechanism as in the case of bosons, the physical foundation of the jump operators is the Pauli blocking (see Fig. \ref{Fig1_Jump_Operators}), i.e. for fermions the spontaneous emission to an already occupied state is blocked \cite{Weimer10}. This gives rise to a novel non-equilibrium pairing mechanism for fermions that does not require attractive conservative forces.

We are primarily interested in states with a homogeneous product of $N$ identical fermion pairs on a two-dimensional (2-D) square lattice:
\begin{equation}
|\Psi\rangle=\eta^{\dag\, N}|{\rm vac}\rangle, \quad \eta^\dag =  \sum_{a,b} \eta_{a,b} c_a^\dag c_b^\dag.
\label{realspacepair}
\end{equation}
Here, $c_a^\dag$ creates a fermion in mode $a$, where $a = (\sigma, i)$ labels spin $\sigma$ and position $i$ on the 2-D lattice. Typically, when we consider delocalised states, the sum extends over the whole lattice in the position space index. For large systems in the thermodynamic limit, we may adopt the grand canonical ensemble, and the state above becomes the BCS-type coherent state for paired fermions:
\begin{equation}
|\Psi\rangle=\mathcal{N}\exp (\sum_{A,B}f_{A,B} c_{A}^{\dag}c_{B}^{\dag})|{\rm vac}\rangle
=\mathcal{N} \prod_{A,B}(1+f_{A,B}c_{A}^{\dag}c_{B}^{\dag})|{\rm vac}\rangle,
\label{bcspairingstate}
\end{equation}
where we have Fourier transformed Eq. (\ref{realspacepair}) into momentum space, with $A=\{\sigma,\mathbf{k}\}$ labeling spin $\sigma$ and momentum $\mathbf{k}$, $\mathcal{N}$ being the normalization factor. The pair wavefunction in momentum space is $f_{A,B}=\sum_{i,j}\eta_{a,b}e^{i\mathbf{k}\mathbf{r}_i+i\mathbf{k}'\mathbf{r}_j}$, where the spin degrees of freedom are not changed. The symmetry of the pairing state is encoded in the form of the pair wavefunction $\eta_{a,b}$ ($f_{A,B}$).

In general, the jump operator that drives the system into the pairing state of Eq. (\ref{bcspairingstate}) can involve multi-particle processes. Single-particle jump operators (involving at most one annihilation mode operator), which act directly on one particle at a time, are generally easier to implement experimentally than two-particle jump operators (involving two annihilation mode operators). However, the two-particle jump operators are the most intuitive, as the dissipative dynamics can be viewed as local phase-locking between pairs of fermions. We will therefore first give a brief description on the derivation of two-particle jump operators for the pairing states, which will provide important clues as to how to proceed with the more practical single-particle operators. In both cases, our focus will be on the d-wave pairing state, though the procedures can be extended to other symmetries as well.

\begin{figure}[tb]
\centerline{\includegraphics[width=8cm]{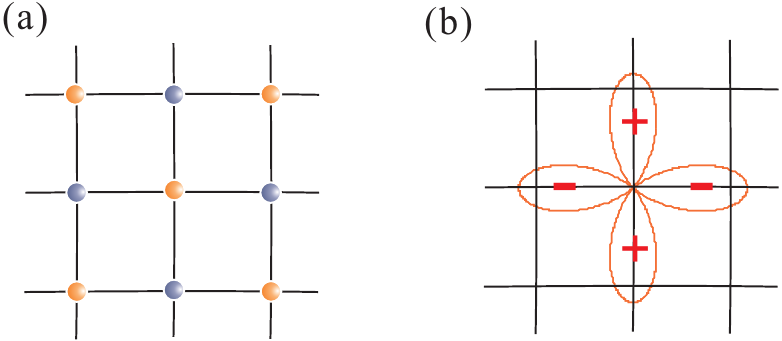}}
\caption{Schematics for the symmetry of antiferromagnetic N\'{e}el state and d-wave state. (a) Spin configurations for N\'{e}el state. Particles with different colors represent different spins; (b) d-wave symmetry on a lattice. Each leaf for the clover-leaf structure here represents singlet pairing of spins on the adjacent sites. The singlet pairs change sign when rotated by $\pi/4$, as dictated by the d-wave symmetry.}
\label{Fig2}
\end{figure}

The fixed-number d-wave pairing state is given by the symmetric superposition of d-wave pairs, which are spin-singlet fermion pairs on the bonds in $\mathbf{x}$ and $\mathbf{y}$ direction whose relative phase changes sign under a $\pi/4$ rotation $\mathbf{x}\leftrightarrow \mathbf{y}$,
\begin{eqnarray}\label{realspacepaird}
|\Psi\rangle_d &=& D^{\dag N}|{\rm vac}\rangle, \\
D^{\dag}&=&  \sum_{j}\left[(c^\dag_{j, \uparrow}  c^\dag_{j+ e_x, \downarrow} - c^\dag_{j,\downarrow}  c^\dag_{j+ e_x,\uparrow})
-(c^\dag_{j, \uparrow}  c^\dag_{j+ e_y, \downarrow} - c^\dag_{j,\downarrow}  c^\dag_{j+ e_y,\uparrow})\right]
\end{eqnarray}
with $e_x,e_y$ the unit lattice vectors long the $\mathbf{x}$ and $\mathbf{y}$ direction, respectively.

Following Eq. (\ref{bcspairingstate}), it is
easy to write down the general form of a fixed-phase d-wave pairing state:
\begin{equation}\label{bcspairingstated}
|\psi\rangle_d=\mathcal{N}\sum_{n}\frac{\alpha^nD^{\dag n}}{n!}|{\rm vac}\rangle=\mathcal{N}\exp\left(\alpha D^{\dag}\right)|{\rm vac}\rangle,
\end{equation}
where $\alpha=e^{i\theta}|\alpha|$ is a complex number carrying the phase $\theta$ of the pairing state. In the following, we will first focus on the fixed-number state and will come back to the fixed-phase state later.

Assuming translational invariance, which is the case for an infinitely large homogeneous lattice or for a finite homogeneous lattice with periodic boundary conditions, we may rewrite the pair creation operators:
\begin{eqnarray}
d_j^\dag = \{(c_{j+e_x\uparrow}^{\dag}+ c_{j-e_x\uparrow}^{\dag})
-(c_{j+e_y\uparrow}^{\dag}+c_{j-e_y\uparrow}^{\dag})\}c_{j\downarrow}^{\dag},
\end{eqnarray}
so that $D^{\dag}=\sum_{j}d^{\dag}_j$. These operators have the advantage that they are already factorised in the spin degrees of freedom, which allows us to write the pairing jump operators in a similar fashion as well. For convenience,
we adopt a shorthand convention and write
$d_j^{\dag}=\sum_{\nu}\rho_{\nu}c^{\dag}_{j+e_{\nu},\uparrow}c_{j,\downarrow}$, where $\rho_{\pm x}=1$, $\rho_{\pm y}=-1$. We also find it useful to define the singlet pairing state in one dimension (1-D):
\begin{equation}
(\eta^{\dag})^N|{\rm vac}\rangle=\left[\sum_j(c^{\dag}_{j\uparrow}c^{\dag}_{j+1\downarrow}+c^{\dag}_{j+1\uparrow}c^{\dag}_{j\downarrow})\right]^N|{\rm vac}\rangle,\label{singletpairing}
\end{equation}
where $c^{\dag}_{i\{\uparrow,\downarrow\}}$ is the single fermion annihilation operator on the
$i$th site with a certain spin. The 1-D state $\left(\eta^{\dag}\right)^N|{\rm vac}\rangle$ captures the off-site singlet pairing feature of the d-wave state. Similar to the d-wave case in 2-D, we may use translational symmetry to define the 1-D singlet pairing operator $\eta^{\dag}_i$:
\begin{equation}
\eta^{\dag}_j=c^{\dag}_{j\uparrow}c^{\dag}_{j+1\downarrow}+c^{\dag}_{j\uparrow}c^{\dag}_{j-1\downarrow}
=c^{\dag}_{j\uparrow}(c^{\dag}_{j+1\downarrow}+c^{\dag}_{j-1\downarrow}),
\end{equation}
so that $\eta^{\dag}=\sum_j\eta^{\dag}_j$.

In the following, we proceed by devising the jump operators for the 1-D singlet pairing state first before considering the 2-D scenarios where additional spatial phase locking is needed.

\subsection{Two-particle jump operators}

We are interested in jump operators that conserve particle number. This implies that they must carry total charge 0 (or with total global phase 0). Thus, we can write it in normal order as a product of a pure creation  and a pure annihilation part. The most general form of the two-particle jump operator thus reads
\begin{equation}
J_i=\chi_i^\dag \xi_i, \quad  \chi_i^\dag = \sum_{a,b} \chi^*_{a,b}(i) c^\dag_a c^\dag_b,\quad \xi_i = \sum_{a,b} \xi_{a,b}(i) c_a c_b.
\end{equation}
Here we also impose the requirement of quasi-locality, i.e. the functions $\chi_{ab}(i), \xi_{ab}(i)$ shall be non-zero only in a small vicinity of site $i$ in position space. Thus the jump operator $J_i$ is centered around site $i$.

To uniquely drive the system into the desired pairing state, it is necessary that the state $|\Psi\rangle$ be a dark state of a set of jump operators $\{\chi_i^\dag \xi_i\}$
\begin{equation}
J_i (\eta^{\dag})^N|{\rm vac}\rangle=\chi_i^\dag \xi_i (\eta^{\dag})^N|{\rm vac}\rangle=0, \quad\forall i.\label{twoparticlejump}
\end{equation}
For a given operator $\xi_i$, we can work out its commutation relation with the creation operator
of the pairs:
\begin{equation}
[\xi_i,\eta^{\dag}]=A, \hspace{2cm} [A,\eta^{\dag}]=B.
\end{equation}

While $A$ carries charge 0 and is composed of a constant plus a normal-ordered second order term, $B$ is a superposition of pair creation operators and carries charge $2$, which implies that $[B,\eta^{\dag}]=0$ (cf. \ref{appendnew}).

With these relations, we find that the commutator of $\xi_i$ with the homogeneous product $\left(\eta^{\dag}\right)^N$ is characterised by the commutators $A$ and $B$ only:
\begin{equation}
\xi_i(\eta^{\dag})^N= (\eta^{\dag})^N\xi_i+ N(\eta^{\dag})^{N-1}A+\frac{N(N-1)}{2}B(\eta^{\dag})^{N-2}\label{gencomm}.
\end{equation}
The first term on the right-hand side of the equation above gives zero when acting on the vacuum. Similarly, the normal-ordered part of $A$ yields zero on the vacuum. Therefore, in order to satisfy the dark state condition, we need to find a set of quasi-local bilinear operators $\xi_i$ and $\chi_i^\dag$ which for a given $\eta^\dag$ uniquely solve the two equations
\begin{equation}
A |{\rm vac}\rangle = 0, \quad \chi_i^\dag B=0.
\label{cond}
\end{equation}

For the 1-D case, after some derivation following the arguments leading to Eq. (\ref{cond}) (see \ref{appendnew} for details), we find a two-particle jump operator of the form:
\begin{equation}
J_i=C^\dag  M c_{i\uparrow}, \quad M = (c^{\dag}_{i+1\downarrow}+c^{\dag}_{i-
1\downarrow})(c_{i+1\downarrow}-c_{i-1\downarrow}), \label{jump}
\end{equation}
where $C^{\dag}$ is an arbitrary superposition of single-fermion creation operators with spin-up. Note that as a spin flip operation only changes the overall sign,  jump operators similar to $J_i$ but with spins flipped also have the singlet pairing state Eq. (\ref{singletpairing}) as a dark state. This is also true for the jump operators of the pairing states that we consider in this work, as similar symmetries in the spin degrees of freedom hold for all the pairing states that we will consider.

For the 2-D case, we find the jump operator (see \ref{appendnew} for details):
\begin{eqnarray}
J_i=C^\dag  M c_{i\uparrow},\quad M = -\sum_{\nu}\rho_{\nu}c^{\dag}_{i+e_{\nu},\downarrow}(c_{i+e_x\downarrow}+c_{i+e_y\downarrow}),
\label{twoparticle2d}
\end{eqnarray}
where $\rho_{\pm x}=1$, $\rho_{\pm y}=-1$. This operator has an interesting structure. It may be seen as a \emph{conditional} dissipative process. The annihilation of a fermion sitting at site $i$ must only take place if a superposition of fermions located at sites $i+ e_x,i+e_y$ is coherently transferred to a superposition on the four sites $\{i\pm e_x,i\pm e_y\}$ centered around site $i$.

The two-particle jump operators above have the desired pairing states as the dark state. However, numerical simulation shows that in most cases they do not guarantee a unique dark state. Furthermore, these jump operators involve correlated dissipation of two particles and are therefore difficult to implement experimentally. Nevertheless the construction scheme above provides clues for the design of jump operators that are easier to implement and with a unique dark state.

\subsection{Single-particle jump operators}

In this subsection, we will show that counter-intuitively, it is  possible to design a set of jump operators for which only single particle operations are required. The single-particle jump operators dissipatively drive fermions into pairs as well as phase lock the pairs into the desired symmetry.
More importantly, we will give arguments later that the dark state of these single-particle jump operators should be unique, and they are easier to implement than the two-particle jump operators. Below, we will describe two ways in which the appropriate single-particle jump operators can be derived.

Firstly, consistent with the discussion in the previous section, in the case of a single-particle jump operator, the annihilation part $\xi_i$ contains a single annihilation operator which must be either $c_{i,\uparrow}$ or $c_{i,\downarrow}$. Hence the operator $A$ carries charge $1$, and we have $B=[A,\eta^{\dag}]=0$. Taking the 1-D singlet pairing state as an example, it is easy to derive that $A=c^{\dag}_{i+1,\downarrow}+c^{\dag}_{i-1,\downarrow}$ for $\xi_i=c_{i,\uparrow}$; and $A=c^{\dag}_{i+1,\uparrow}+c^{\dag}_{i-1,\uparrow}$ for $\xi_i=c_{i,\downarrow}$. Thus to satisfy Eq. (\ref{twoparticlejump}), we need $\chi_i A=0$, and the simplest choice is $\chi_i^{\dag}=(c^{\dag}_{i+1,\downarrow}+c^{\dag}_{i-1,\downarrow})$ for $\xi_i=c_{i,\uparrow}$; and  $\chi_i^{\dag}=(c^{\dag}_{i+1,\uparrow}+c^{\dag}_{i-1,\uparrow})$ for $\xi_i=c_{i,\downarrow}$. Considering the superpositions of these operators, we have the set of jump operators:
\begin{equation}
J_{i}^{a}=(c^{\dag}_{i+1}+c^{\dag}_{i-1})\sigma^{a}c_i, \label{singlejump}
\end{equation}
with two-spinor $c_i=(c_{i,\uparrow},c_{i,\downarrow})^T$ and Pauli matrices $\sigma^a$ with $a=\pm,z$.

Alternatively, we start with the physical intuition that d-wave pairing states may be viewed as delocalised antiferromagnetic order away from half-filling. We may consider a unit cell of the N\'{e}el state $S_{i,j}^{\pm}|{\rm vac}\rangle$, where the N\'{e}el state unit cell operator $S^{\pm}_{i,j}=c^{\dag}_{i}\sigma^{\pm}c^{\dag}_{j}$, and $j$ is one of the nearest neighbors of site $i$ which thus creates an adjacent pair of fermions with opposite spin. We notice that the singlet pairing operator in 1-D can be viewed as the superposition of two antiferromagnetic unit cell operators $\eta^{\dag}_i=S_{i,i+1}^{+}+S_{i,i-1}^{+}$. It is thus instructive to first construct the jump operators for an antiferromagnetic N\'{e}el state. To annihilate this unit cell N\'{e}el state, we simply need the Lindblad operators of the form:
\begin{equation}
j_{i,j}^{\pm}=c_{i}^{\dag} \sigma^{\pm}c_{j}.\label{1dafjump}
\end{equation}
This jump operator generates hopping with a spin flip, which is impossible in the case that antiferromagnetic order is already present, due to the Fermi statistics. Generalising the unit cell operator to a 2-D lattice, we notice the N\'{e}el state can be written in eight
different forms, $|{\rm N}\pm\rangle=\prod_{i\in A}S_{i,i+\mathbf{e}_{\nu}}^{\pm}|{\rm vac}\rangle=(-1)^{M/2}\prod_{i\in B}S_{i,i-\mathbf{e}_{\nu}}^{\mp}|{\rm vac}\rangle$, with $M$ the lattice size, and $\mathbf{e}_\nu=\{\pm\mathbf{e}_x,\pm\mathbf{e}_y\}$. The 2-D Lindblad operators corresponding to those in Eq. (\ref{1dafjump}):
\begin{eqnarray}\label{AFjump}
j_{i,i+\mathbf{e}_{\nu}}^{a}=c_{i+\mathbf{e}_{\nu}}^{\dag} \sigma^{a}c_{i}, \, i \in A\,{\rm or}\, B.
\end{eqnarray}
These operators impose the quasi-local constraint on the steady state that any given site should have opposite spin with its nearest neighboring sites, which guarantees antiferromagnetic order at half-filling. However, there is still a two-fold degeneracy where the antiferromagnetic order differs by a total spin flip. However, there is still a two-fold degeneracy where the antiferromagnetic order differs by a total spin flip. As the N\'{e}el states cannot be reached by implementing the jump operators in Eq. (\label{AFjump}) on any other states, and as the complete set of jump operators is invariant under a Hermitian conjugation, the dark subspace, (containing the N\'{e}el states) is isolated from the rest of the Hilbert space under the dissipative dynamics given by Eq. (\label{AFjump}). These problems can be solved, in principle, by adding a single jump operator, or more naturally a set of jump operators of the form:
\begin{equation}
j_i=c^{\dag}_{i,\sigma}c_{j,\sigma},\label{afjump2}
\end{equation}
for arbitrary $i$ and its nearest neighbor $j$, and for either spin $\sigma=\uparrow,\downarrow$. As an example, for $i\in A$ and $\sigma=\downarrow$, the dark state is $|{\rm N}+\rangle$.  Note that comparing the jump operators in Eq. (\ref{1dafjump}) with the unit cell operators $S^{\pm}_{i,j}$, we find that the jump operators can be obtained from the state generating unit cell operator via a particle-hole transformation $c_j^{\dag}\rightarrow c_j$ on the central site $j$.

Now that we have found the Lindblad operators for the N\'{e}el state, we can proceed to generalise the operators to the d-wave case. First, we identify the ``d-wave unit cell operators":
\begin{equation}\label{dunitcell}
\hat{D}_{i}^{a}  =\sum_{\nu}\rho_\nu S_{i,i+\mathbf{e}_\nu}^{a},
\end{equation}
where $\rho_{\pm x}=-\rho_{\pm y}=1$ and $a=\pm$. We then perform a particle-hole transformation on the central site as in the case of the N\'{e}el state, and find operators:
 \begin{equation}
J_{i}^{a}=\sum_{\nu}\rho_\nu j_{i,i+\mathbf{e}_\nu}%
^{a},\,\,J_{i}^{z}=\sum_{\nu}\rho_\nu j_{i,i+\mathbf{e}_\nu}^{z}.
\label{DWjump}
\end{equation}
It is easy to verify that $[J_{i}^{\alpha},\sum_{j}\hat{D}_{j}^{b}]=0$ ($\alpha=\pm,z)$, which is dictated by the Fermi statistics. Note that the d-wave coherence is established via quasi-local phase locking between adjacent cloverleaves of sites, as illustrated in Fig. \ref{Fig2}. In this respect, the jump operators act similarly to those that establish phase coherence in bosonic systems \cite{Diehl08}. For fermionic pairing states, however, as explained above, Pauli blocking is an additional key ingredient.

We now consider the general case of quasi-local pairing states that can be factorised in the spin degrees of freedom. The unit cell operator of these pairing states can be expressed as superpositions of the N\'{e}el state unit cell operators:
\begin{equation}
\beta^{\dag}_i=\sum_{\nu}\rho_{\nu}c^{\dag}_{i+\mathbf{e}_\nu,\sigma_1}c^{\dag}_{i,\sigma_2},
\end{equation}
where the coefficients $\rho_{\nu}$ encode the spatial symmetry of the pairs, and $\sigma_1$ and $\sigma_2$ can be arbitrary combinations of spin configurations, including spinless fermions $\sigma_1=\sigma_2$. Performing the particle-hole transformation as above, we get the following jump operators:
\begin{equation}
J_i=\sum_{\nu}\rho_{\nu}c^{\dag}_{i+\mathbf{e}_\nu,\sigma_1}c_{i,\sigma_2},
\end{equation}
which gives $[J_j,\sum_i\beta^{\dag}_i]=0$, so long as the coefficients satisfy the following equations:
\begin{eqnarray}
\sum_{\mu,\nu}\rho_{\mu}\rho_{\nu}c^{\dag}_{j+\mathbf{e}_\nu,\sigma_1}c^{\dag}_{j+\mathbf{e}_\mu,\sigma_1}=0,\quad&&{\rm for }\hspace{0.2cm}\sigma_1\neq \sigma_2,\label{zeroeqn}\\
\sum_{\mu,\nu}\rho_{\mu}\rho_{\nu}(c^{\dag}_{j+\mathbf{e}_\nu}-c^{\dag}_{j-\mathbf{e}_\nu})c^{\dag}_{j+\mathbf{e}_\mu}=0,\quad&&{\rm for }\hspace{0.2cm}\sigma_1=\sigma_2,\label{spinless}
\end{eqnarray}
where the spin degrees of freedom have been factored out, and we have omitted the spin index in Eq. (\ref{spinless}), as there is only one spin species.

For the spinful case, it is easy to verify that Eq. (\ref{zeroeqn}) always holds regardless of the structure of $\rho_{\nu}$. The spinless case though, is non-trivial in general. A particularly interesting example following Eq. (\ref{spinless}) above is a 2-D p-wave state of spinless fermions generated by $p^\dag \sim \sum_{i,\nu}\rho_\nu c^\dag_{i+\mathbf{e}_\nu}c^\dag_{i}$ with $\rho_{x}=-\rho_{-x}=-\mathrm i\rho_{y}= \mathrm i \rho_{-y} =1$, the Lindblad operators are $\{\sum_\nu \rho_\nu c^\dag_{i+\mathbf{e}_\nu}c_{i}\}$.  The p-wave pairing state in 2-D can be prepared in two different chiralities (see Fig. \label{Figpwave}): $p_x+ip_y$ and $p_x-ip_y$, which shares the spatial symmetry with the pairing states in topological superconductors. However, as we will show later in the mean-field analysis, the p-wave states prepared in this way are still in the strong pairing limit, such that they are topologically trivial. The generation of topological order in 2-D systems will be discussed in a forthcoming publication. For stable dissipatively induced topological order in 1-D, see \cite{Diehl11}.

\begin{figure}[tb]
\centerline{\includegraphics[width=8cm]{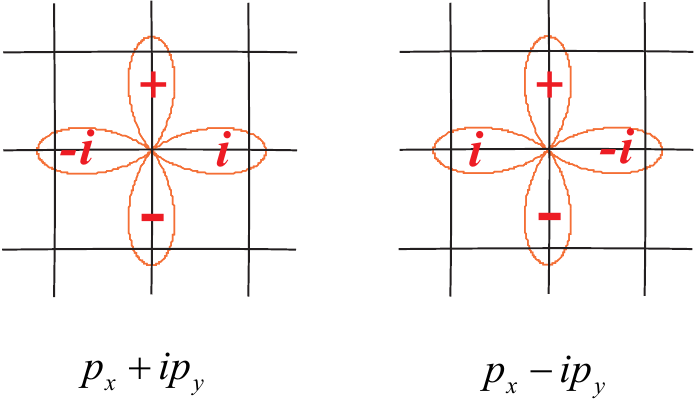}}
\caption{Schematics for the p-wave symmetry on a lattice for spinless fermions. Each leaf for the clover-leaf structure stands for triplet pairing of particles on the adjacent sites. The overall phase of the triplet pairs change by $i$ or $-i$ under a $\pi/4$ rotation, as dictated by the p-wave symmetry.}
\label{Figpwave}
\end{figure}

\subsection{Jump operators for fixed-phase state}\label{fixphasesec}

In the previous discussion, we have focused on the jump operators for dark states in the form of Eq. (\ref{realspacepaird}), i.e. states with fixed total particle number. The dissipative processes characterised by these jump operators necessarily conserve the total particle number of the system, as is clear from the commutation relation $[J_i^{a,z},\hat{N}]=0$, where $\hat{N}$ is the total particle operator. The total particle number of the dark state in this case is given by that of the initial state.

On the other hand, one can show (c.f. \ref{appendixb}) that for any given number-conserving pairing state which is a dark state of single-particle jump operators, one can always construct a set of linear jump operators that have the fixed-phase state in the form of Eq. (\ref{bcspairingstated}) as a dark state. In the case of the d-wave pairing state, we consider the following jump operators
\begin{eqnarray}
j_{i,\uparrow}&=&-P\sum_{\nu}\rho_{\nu}c^{\dag}_{i-\mathbf{e}_{\nu},\downarrow}+Qc_{i,\uparrow},\\
j_{i,\downarrow}&=&P\sum_{\nu}\rho_{\nu}c^{\dag}_{i+\mathbf{e}_{\nu},\uparrow}+Qc_{i,\downarrow},
\end{eqnarray}
where $\rho_{\pm x}=-\rho_{\pm y}=1$ as given by the d-wave symmetry, and $P$ and $Q$ are complex numbers. It is easy to show that $j_{i,\sigma}|\psi\rangle_d=0$, provided that $P/Q=\alpha$, with the fixed-phase pairing state $|\psi\rangle_d$ defined in Eq. (\ref{bcspairingstated}). Note that the jump operators for the fixed-phase state do not conserve the total particle number, while the average particle density in the dark state is given as $N=\sum_{\mathbf{q},\sigma}\left\langle\psi|c^{\dag}_{\mathbf{q},\sigma}c_{\mathbf{q},\sigma} |\psi\right\rangle_d=\sum_{\mathbf{q}}\frac{2|\alpha\varphi(\mathbf{q})|^2}{1+|\alpha\varphi(\mathbf{q})|^2}$, where $\varphi(\mathbf{q})$ is the pair wavefunction in momentum space, and the summation over $\mathbf{q}$ runs over the first Brillouin zone. Importantly, the average particle density of the dark state here is determined by the parameters of the dissipative process, i.e. $|\alpha|$ can be chosen to fix a desired average density.

It is straightforward to extend the analysis to pairing states with other symmetries. For the spinless p-wave pairing state for example, the jump operators for the fixed-phase dark state $|\psi\rangle_p=\mathcal{N}\exp(\alpha p^{\dag})|{\rm vac}\rangle$ are
\begin{equation}
j_i=P\sum_{\nu}\rho_{\nu}c^{\dag}_{i+\mathbf{e}_{\nu}}+Qc_{i},
\end{equation}
with $P/Q=2\alpha$, and the factor of $2$ is due to the triplet pairing symmetry of the p-wave pairing state. For a more detailed discussion on jump operators for fixed-phase states, see \ref{appendixb}.

\section{Completeness of the jump operators} \label{symmetryanalysis}

To achieve the dissipative preparation of many-body states, the final steady state of the master equation should be unique. This requires that (i) the dark state be unique; (ii) the non-existence of stationary solutions other than this dark state \cite{Kraus}. Taking the 1-D singlet pairing state as an example, we shall first analyse the uniqueness of the dark state from the symmetry perspective. We will then illustrate the uniqueness of the steady state with different pairing symmetries both in 1-D and 2-D by directly evolving the master equation for small finite size systems.

\subsection{Uniqueness of dark state from a symmetry perspective}
The problem of showing the uniqueness for all Lindblad operators is equivalent to showing the uniqueness of the ground state of the following positive semi-definite Hamiltonian (dimensionless):
\begin{eqnarray}
H_p=\sum_{i,\alpha} (J_{i}^{\alpha})^{\dag} J_{i}^{\alpha}, \label{parentH}
\end{eqnarray}
where $J_{i}^{\alpha}$ ($\alpha = \pm,z$ or $x,y,z$) are the d-wave jump operators in Eq. (\ref{DWjump}). To see the equivalence, note that for any operator $A$, the matrix element $\langle \psi |A^\dag A |\psi\rangle \geq 0$  is non-negative. Thus, the spectrum of the Hermitian Hamiltonian is real and non-negative. Any zero energy eigenstate must be a ground state of the Hamiltonian. The energy is zero if and only if each term has zero energy. This is equivalent to $J_{i}^\alpha|\Psi\rangle =0 \,\forall \, i,\alpha$. Hamiltonians with these properties often occur in the context of spin models, where they are constructed as ``parent Hamiltonians'' for given states \cite{auerbachbook}. We follow this nomenclature here. This Hamiltonian serves for the uniqueness considerations of this section as well as for the adiabatic passage discussed in Sec. \ref{adHubbard}.

For later convenience, it is useful to collect the $\alpha = \pm$ components into a dimensionless ``reduced'' parent Hamiltonian, with the normal-ordered form
\begin{eqnarray}
H_p^{r}=-\sum_{i,\sigma} (c^{\dag}_{i+1,\sigma}+c^{\dag}_{i-1,\sigma})c^{\dag}_{i,-\sigma}c_{i,-\sigma}(c_{i+1,\sigma}+c_{i-1,\sigma})
+2\sum_{i,\sigma}c^{\dag}_{i,\sigma}c_{i,\sigma},
\label{redparentH}
\end{eqnarray}
where $\sigma=\uparrow,\downarrow$, and the quartic terms describe effective attractive two-body interactions. Note that the ``chemical potential'' term proportional to the total particle number is unimportant for fixed particle number states.
For $\alpha=z$, the normal-ordered form (dimensionless) is
\begin{eqnarray}
H_p^{z}&=&\sum_{i,\sigma}(c^{\dag}_{i+1,\sigma}+c^{\dag}_{i-1,\sigma})c^{\dag}_{i,-\sigma}c_{i,\sigma}(c_{i+1,-\sigma}+c_{i-1,-\sigma})\nonumber\\
&-&\sum_{i,\sigma} (c^{\dag}_{i+1,\sigma}+c^{\dag}_{i-1,\sigma})c^{\dag}_{i,\sigma}c_{i,\sigma}(c_{i+1,\sigma}+c_{i-1,\sigma})
+2\sum_{i,\sigma}c^{\dag}_{i,\sigma}c_{i,\sigma},
\label{zparentH}
\end{eqnarray}
where the first term describes correlated hopping, and we have $H_p=H_p^{r}+H_p^z$. Following the discussion in the previous paragraph, we see that the d-wave state $|\Psi\rangle_d$ is a ground state of the full parent Hamiltonian as well as of the reduced parent Hamiltonian. We will further demonstrate below that while the d-wave state is not the unique ground state of the reduced parent Hamiltonian, there are strong indications for it to be the unique ground state for the complete parent Hamiltonian based on symmetry arguments. These considerations will play an important role later in designing the implementation schemes.

By construction of the jump operators, the $d$-wave states are ground states of the parent Hamiltonian. For any symmetry operation, i.e. a unitary transformation $T$, that leaves the parent Hamiltonian Eq. (\ref{parentH}) invariant $T H_p T^{-1} = H_p$, a necessary but not sufficient condition for the d-wave state to be the unique ground state of the parent Hamiltonian is:
\begin{eqnarray}
T |\Psi \rangle_d = \exp\left( i \phi_T\right)|\Psi \rangle_d, \label{symmT}
\end{eqnarray}
where $\phi_T$ is the phase imposed onto the d-wave state by the unitary operation $T$.

The parent Hamiltonian Eq. (\ref{parentH}) has two obvious global symmetries: phase rotation invariance associated with particle number conservation, and translational invariance associated with momentum conservation. In the following, we discuss these symmetries:

\emph{Phase rotation invariance} -- The symmetry is generated by $T_\varphi= \exp \mathrm i \varphi \hat N$, where the number operator $\hat N = \sum_{i,\sigma}  c_{i,\sigma}^\dag c_{i,\sigma}$. Since the d-wave state is an eigenstate for both $H_p$ and $\hat N$, with $H_p|\Psi\rangle_d=0$ and $\hat N|\Psi\rangle_d=2N$, it is also an eigenstate of $T_\varphi$ with eigenvalue $\exp 2\mathrm i \varphi N$. Thus for a given fixed particle number no degeneracies occur according to the above criterion.

\emph{Translation invariance} -- The symmetry is generated by the total center-of-mass momentum operator in 2-D:
\begin{eqnarray}
\hat {\vec{P}} &=& \frac{\mathrm i}{2} \sum_{i,\sigma}\{ ( c_{i+e_x,\sigma}^\dag - c_{i-e_x,\sigma}^\dag)c_{i,\sigma} \vec{e}_{x}
+( c_{i+e_y,\sigma}^\dag - c_{i-e_y,\sigma}^\dag)c_{i,\sigma}\vec{e}_y\}\\
T_{\mathbf{r}} &=&  \exp \mathrm{i} \mathbf{r} \hat{\vec{P}},
\end{eqnarray}
for which $T_{\mathbf{r}}H_p T_{\mathbf{r}}^{-1}=H_p$. It is easy to check that $\hat{\vec{P}} |\Psi\rangle_d =0$, i.e. the $d$-wave state is a momentum eigenstate with zero eigenvalue. Therefore, the translational symmetry does not lead to degeneracies. Note the close relation of the terms in the momentum operator to the jump operators, which may be transformed into each other by flipping the spin on the site in the middle and changing from the symmetric to the antisymmetric superposition on sites $\{i\pm e_x,i\pm e_y\}$.

\emph{Discrete symmetries} -- On the bipartite lattices, and only for those, we find an additional discrete symmetry for the reduced parent Hamiltonian Eq. (\ref{redparentH}), but not the complete parent Hamiltonian Eq. (\ref{parentH}). A bipartite lattice can be split into two equivalent sublattices ($A,B$) in such a way that each lattice site of $A(B)$ is surrounded by lattice sites of $B(A)$. Examples are equally spaced lattices with even number of sites and periodic boundary conditions in one dimension, or even site square lattices with periodic boundary conditions in two dimensions. Given a certain finite lattice in any dimension, it is straightforward to check bipartiteness.

On the bipartite lattices we find the symmetry:
\begin{eqnarray}
T_d: \;\;&&c_{i,\uparrow} \to  -  c_{i,\uparrow};\quad  c_{i,\downarrow} \to c_{i,\downarrow} \quad {\rm for } \hspace{0.2cm}i\in A,\nonumber\\
&&c_{i,\uparrow} \to   c_{i,\uparrow};\;\;\;\;\;\; c_{i,\downarrow} \to c_{i,\downarrow} \quad {\rm for }\hspace{0.2cm} i\in B, \nonumber
\end{eqnarray}
and analogous for $c_{i,\sigma}^\dag$, so that $T_d H_p^r T_d^{-1}=H_p^r$, while $T_d H_p T_d^{-1}\neq H_p$. This transformation is canonical. A second quantised representation of the symmetry is highly non-local, similar to Shiba transformations for the Fermi-Hubbard model \cite{FCZhang}, but in this context we only need its action on the Hamiltonian and the state. Applying the transformation to the $d$-wave state we have
\begin{eqnarray}
T_d|\Psi\rangle_d \equiv \sum_i \left(T_d D_i^{a}\right)^N|{\rm vac}\rangle,\quad T_d D_i^a=\sum_{\nu}\rho_{\nu}f(i,a)S^{a}_{i,i+\mathbf{e}_\nu},
\end{eqnarray}
where $f(i,a)=1$ for $a=+,i\in B$ and for $a=-,i\in A$; $f(i,a)=-1$ for $a=+,i\in A$ and for $a=-,i\in B$. Thus the $d$-wave is not an eigenstate of $T_d$ which implies degeneracy under $H^r_p$ but not $H_p$. The degeneracy emerging from this is two-fold for any lattice size, which we will see in later sections.

In general these symmetries help to classify the lattice configurations (especially for finite lattices) under which we can expect uniqueness. Provided that there are no other symmetries of the full parent Hamiltonian that transform the d-wave state into other distinctive states, the d-wave state should be the unique dark state. Though we cannot rule out constructively the existence of other symmetries which may bring in additional degeneracies, the analysis here provides useful insights on the possible existence of degeneracy.

\begin{figure}[tb]
\includegraphics[width=8cm]{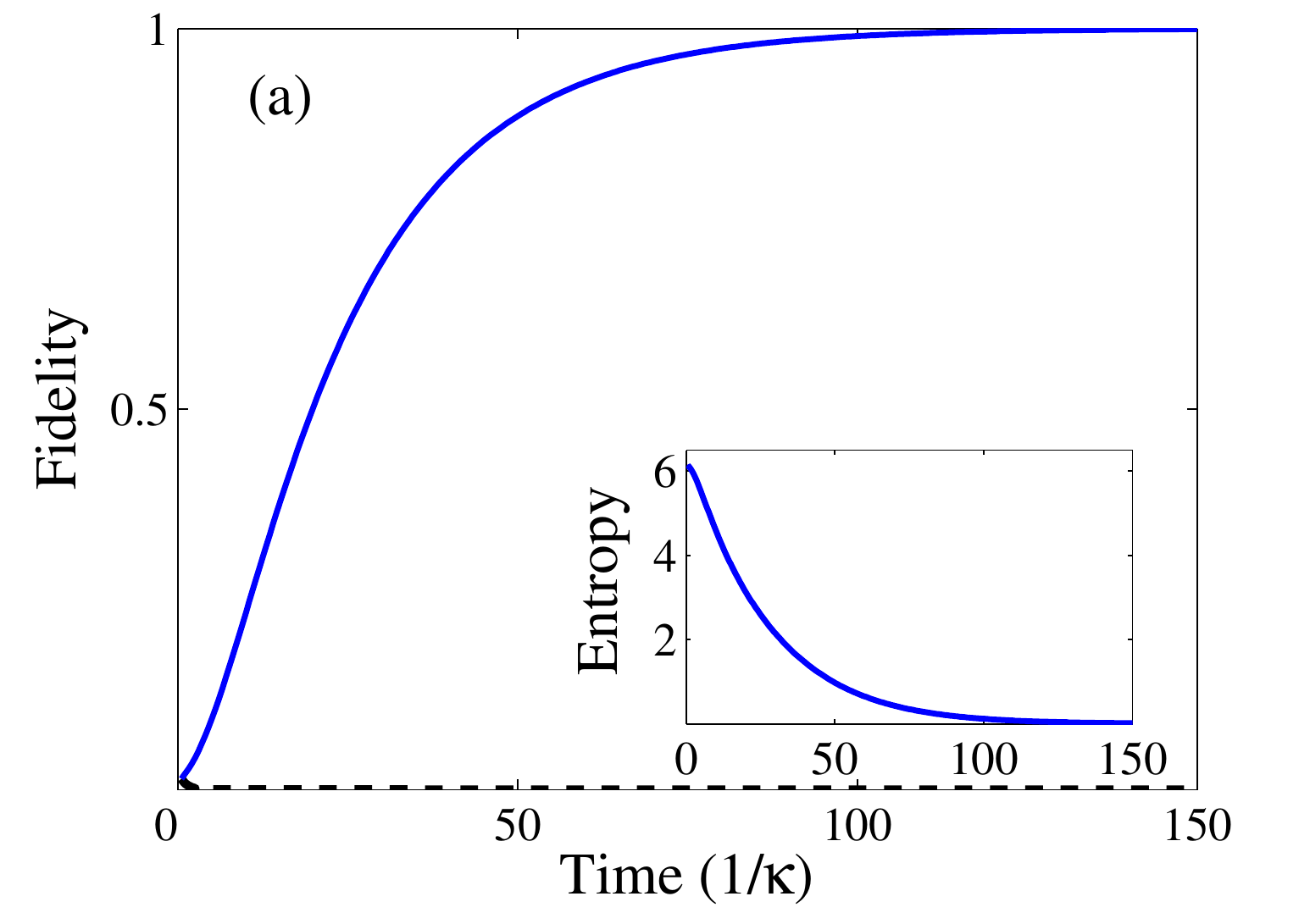}
\includegraphics[width=8cm]{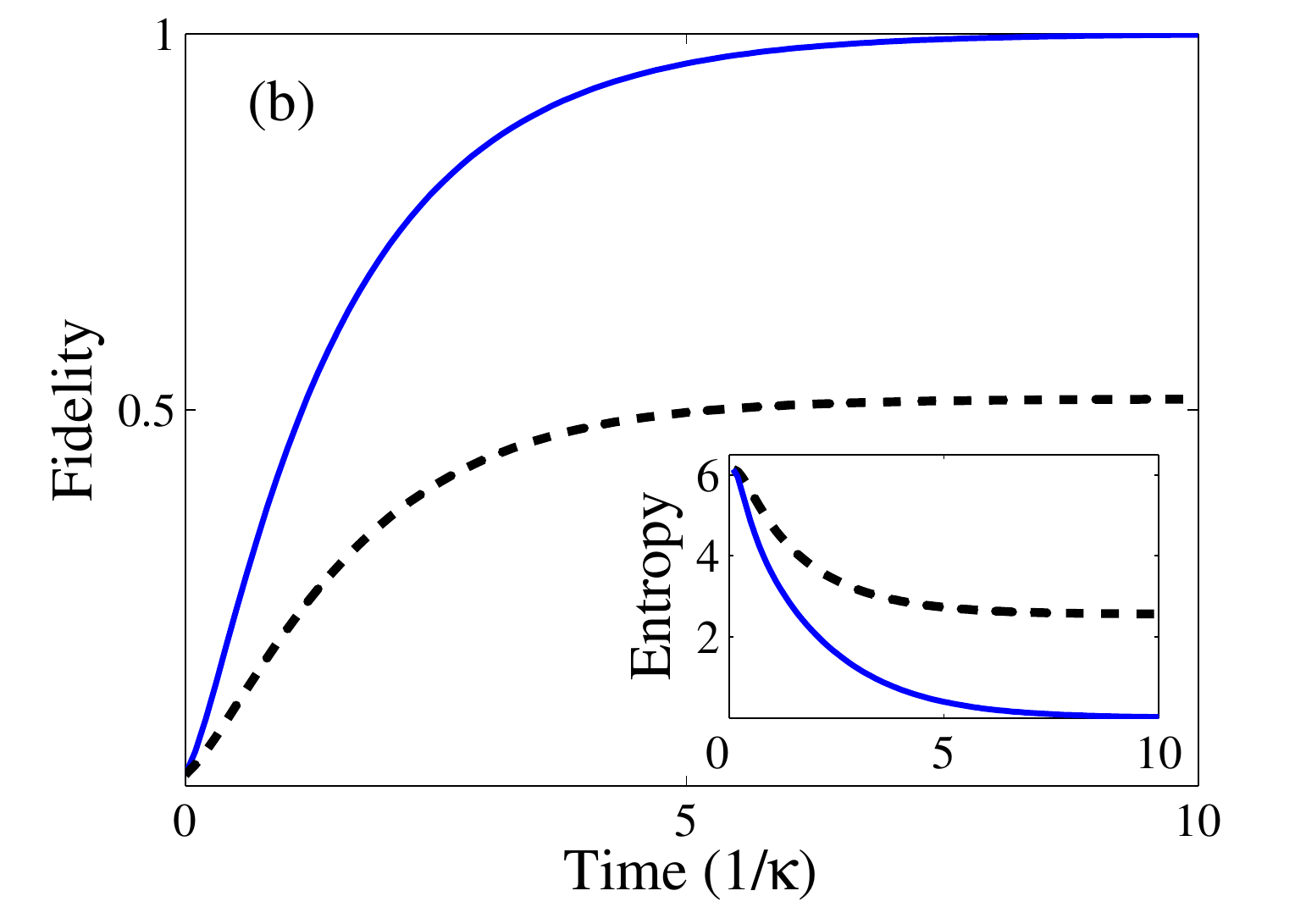}
\caption{Fidelity and entropy evolution of the master equations for 4 atoms on a 1-D chain with 4 sites. (a) The fidelity is with respect to an antiferromagnetic N\'{e}el state. The dashed curve represents the evolution of the fidelity with respect to the other antiferromagnetic state of the system with a total spin flip; (b) The fidelity is with respect to a 1-D singlet pairing state. The dashed curve shows the evolution without $\{J^z_i\}$ jump operators.}
\label{Fig1daf}
\end{figure}

\subsection{Uniqueness of the steady state: Numerical simulations}\label{steadyunique}

To further verify the uniqueness of the dark state, and more importantly, the uniqueness of the steady state for the dissipative dynamics given by the jump operators, we have performed numerical simulations of the master equation dynamics on finite size systems. Due to the translational symmetry of the pairing states, we impose periodic boundary conditions on the finite 1-D and 2-D systems. We have also taken periodic boundary conditions on the jump operators to reduce finite size effects and to be consistent with the definition of the pairing states on a finite lattice. In sufficiently large systems, the jump operators drive the atoms in the bulk into the desired phase, and the mixing of states at the boundary becomes negligible so long as the system is in the thermodynamic limit.

\begin{figure}[tb]
\includegraphics[width=8cm]{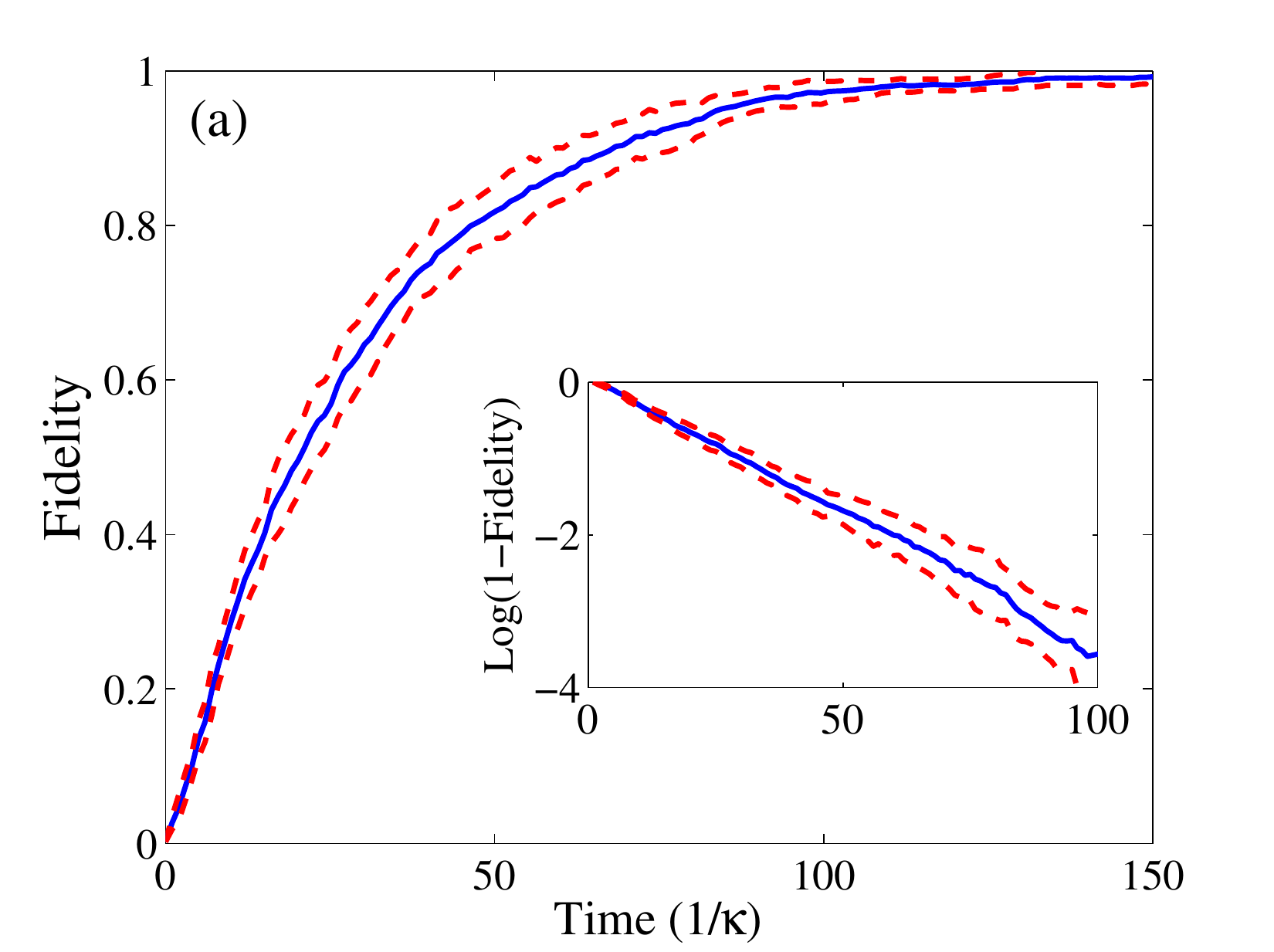}
\includegraphics[width=8cm]{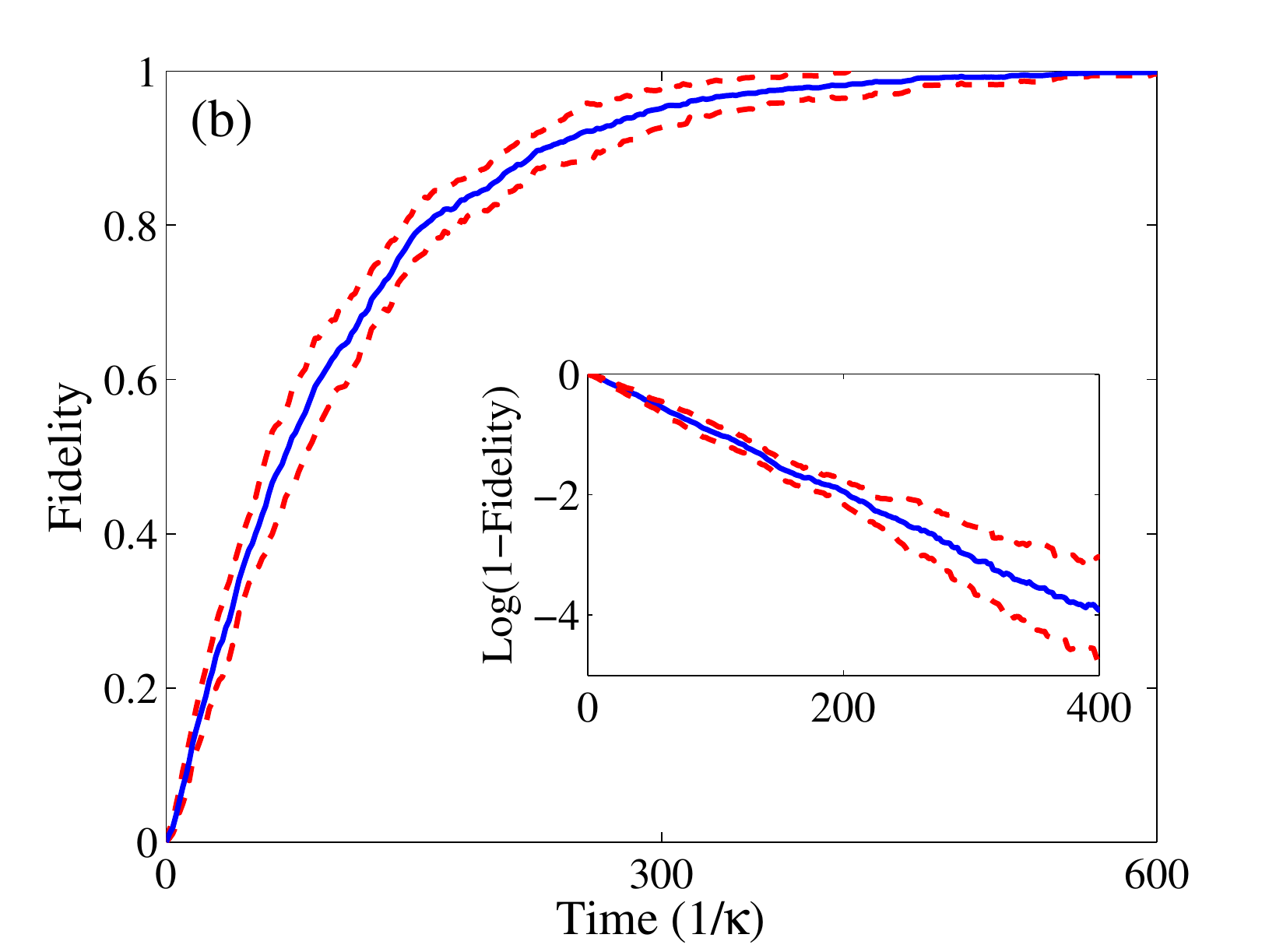}
\caption{Quantum trajectory evolution of the master equations for d-wave and p-wave states in 2 dimensions. (a) Evolution with d-wave jump operators on a 2$\times$6 ladder with 4 atoms; (b) evolution with p-wave jump operators on a 4$\times$4 plaquette with 4 atoms. The insets indicate the existence of dissipative gaps in both cases, which render the convergence to the steady states exponentially fast. This result is robust in the thermodynamic limit as revealed by our mean-field theory. The fidelity (solid) is calculated by averaging over $1000$ trajectories. These trajectories are then bunched into $100$-trajectory groups, whose standard deviations are then calculated to show the sampling errors (dashed).}
\label{Fig2dfid}
\end{figure}

We have evolved the master equations for antiferromagnetic N\'{e}el states and singlet pairing states for finite size systems in 1-D. The results are shown in Fig. \ref{Fig1daf}, which clearly demonstrate the uniqueness of the steady state anticipated above. Both the fidelity and entropy evolution indicates that in both cases the system is driven into the desired state regardless of the initial state. In comparison, when only jump operators of the reduced parent Hamiltonian are applied (dotted curve in Fig. \ref{Fig1daf}(b)), the final fidelity approaches $0.5$. For d-wave pairing states in 2-D, we have carried out a quantum trajectory simulation for small plaquettes. The evolution of the fidelity with respect to the d-wave pairing state indicates that the system approaches the final pure steady state exponentially, which implies the existence of a dissipative gap, analogous to the energy gap for the ground state of the BCS pairing state (see Fig. \ref{Fig2dfid}(a)). While in a finite system, a gap of the order $L^{-2}$ is expected due to the finite linear dimension $L$, in the thermodynamic limit, this gap vanishes. In the next section, we will derive a mean-field theory of the master equation in the thermodynamic limit, which shows that a dissipative gap appears naturally from the mean-field expansion of the master equation for pairing states, demonstrating the dissipative gap in our numerical simulations is not a mere finite size effect. Similar results can be obtained for the paired states of p-wave symmetry for spinless fermions (see Fig. \ref{Fig2dfid}(b)).

\section{Mean-field expansion of the master equation and the dissipative gap}\label{meanfieldsection}

In this section, we will develop a mean-field theory of the master equation for driven-dissipative pairing states in the thermodynamic limit which is valid at late times, where the system is close to the BCS-type pairing state.
For this purpose, we switch from the fixed-number state representation to the fixed-phase (coherent state) representation. The justification for this procedure builds on two properties. First, the exactly known fixed-number pairing dark states discussed above exhibit phase locking among different fermion pairs. Such a property is reproduced in the coherent state representation $|\psi\rangle_d\propto \sum_n\frac{\alpha^n(D^{\dag})^n}{n!}|{\rm vac}\rangle=\exp \left(\alpha D^{\dag}\right)|{\rm vac}\rangle$ ($\alpha=|\alpha|e^{i\theta}$), with a fixed global phase $\theta$ [c.f. Eq. (\ref{bcspairingstated})]. Second, for the BCS wavefunctions under consideration, these representations are equivalent in the thermodynamic limit, which can be verified explicitly from a consideration of the relative number fluctuations in the fixed-phase state. Indeed, one finds for the variance $\Delta N = \left(\frac{\langle \hat N^2 \rangle - \langle \hat N \rangle^2}{\langle \hat N \rangle^2}\right)^{1/2} \sim \frac{1}{\sqrt{N}}$, where $\hat N$ is the total particle number operator, the average is taken with respect to the fixed-phase BCS state, and $N$ is the number of degrees of freedom. The general normalised wavefunction of the BCS pairing state then reads \cite{leggettbook2}:
\begin{equation}
\vert D(\theta)\rangle = \prod_{\mathbf{q}} [\frac{1}{\sqrt{1+ |\alpha\varphi_{\mathbf{q}}|^2} } + \frac{e^{i\theta}|\alpha|\varphi_\mathbf{q}}{\sqrt{1+ |\alpha\varphi_{\mathbf{q}}|^2} } c^\dag_{\mathbf{q},\uparrow} c^\dag_{-\mathbf{q},\downarrow} ] \vert{\rm vac}\rangle,
\end{equation}
where $\varphi_\mathbf{q}$ is the pair wavefunction in momentum space, $\theta$ is the overall phase of the pairing state, $|\alpha|$ is a real number fixing the average particle number density, and the product over $\mathbf{q}$ runs over the first Brillouin zone. Note that for 1-D singlet state, $\varphi_q=\cos q$, while for 2-D d-wave state, $\varphi_{\mathbf{q}}=\cos(q_x)-\cos(q_y)$. The fixed-number d-wave state can be obtained from the coherent state representation via a number projection,
\begin{eqnarray}
 |\Psi\rangle_d = \int\frac{d\theta}{2\pi} e^{\mathrm i \theta \hat {N}}|D(\theta)\rangle = (D^\dag )^{N}|{\rm vac}\rangle,
\end{eqnarray}
where $N$ is the total particle number corresponding to the density
\begin{equation}\label{numeqn}
n=\int \frac{d\mathbf{q}}{(2\pi)^d} \frac{2|\alpha\varphi_{\mathbf{q}}|^2}{1+|\alpha\varphi_{\mathbf{q}}|^2}.
\end{equation}
In the following, we will take $\alpha=1$ for simplicity, which corresponds to an initial state with given total particle density $n=\int \frac{d\mathbf{q}}{(2\pi)^d} \frac{2|\varphi_{\mathbf{q}}|^2}{1+|\varphi_{\mathbf{q}}|^2}$. For initial states with different total particle densities, one just needs to substitute $\varphi_{\mathbf{q}}$ in the following discussions with $\alpha\varphi_{\mathbf{q}}$, where $|\alpha|$ is determined from the number equation Eq. (\ref{numeqn}). Finally, we note that in the grand canonical ensemble, the overall phase $\theta$ of the pairing state can take any value.  Hence we will take $\theta=0$ in the following discussion and write $|D\rangle\equiv|D(0)\rangle$. Indeed, the phase $\theta$ is not determined by the microscopic dynamics, since the jump operators are particle number conserving (charge $0$). Therefore, the phase will be chosen spontaneously in the thermodynamics limit. This is the analog of the spontaneous symmetry breaking in the dissipative context.

The d-wave state has three non-zero bilinear expectation values for each momentum mode relevant to the corresponding quantities in the subsequent mean-field theory, i.e. particle number and order parameter:
\begin{eqnarray}
n_{\mathbf{q}} &=& \langle D \vert c^\dag_{\pm \mathbf{q},\sigma} c_{\pm \mathbf{q},\sigma} \vert D\rangle = \frac{|\varphi_{\mathbf{q}}|^2}{ 1 + |\varphi_{\mathbf{q}}|^2},\label{MFExpVal}\\
\Delta_{\mathbf{q}} &=&\langle D \vert c_{\pm \mathbf{q},\uparrow} c_{\mp \mathbf{q},\downarrow} \vert D\rangle = -  \frac{\varphi_{\mathbf{q}}}{ 1 + |\varphi_{\mathbf{q}}|^2} ,\label{MFExpValgap1}\\
\Delta^{\ast}_{\mathbf{q}} &=&\langle D \vert c^{\dag}_{\mp \mathbf{q},\downarrow} c^{\dag}_{\pm \mathbf{q},\uparrow} \vert D\rangle = -  \frac{\varphi^{\ast}_{\mathbf{q}}}{ 1 + |\varphi_{\mathbf{q}}|^2} ,\label{MFExpValgap2}
\end{eqnarray}
where $\sigma = \uparrow,\downarrow$. All other expectation values vanish on the above state.

A mean-field theory analogous to the BCS approximation for superconductivity can be set up based on the proximity of the density matrix to the d-wave state, giving rise to an ordering principle for devising a controlled mean-field approximation in the late time evolution, which is thus useful to study the final stages of the master equation evolution.

Starting from the fact that the coherent representation of the pairing state is a \emph{product} in momentum space, we will require this property also for an approximate ansatz for the density matrix, $\rho = \prod_{\mathbf{q}} \rho_{\mathbf{q}}$ for the solution of the master equation at late times. This may be viewed as a Gutzwiller factorisation approach in momentum space, which has been used previously for a mean-field decoupling of bosonic master equations \cite{tomadin}. This ansatz will enable us to derive a late time master equation quadratic in the fermion operators which contains information of the complex excitation spectrum, i.e. the damping of the lowest fermionic single particle excitations.

To implement the approximation, we Fourier transform the jump operators to momentum space:
\begin{equation}
J_{k}^{\alpha}=\sum_{\mathbf{q}} \varphi_{\mathbf{q}} c^{\dag}_{\mathbf{q}}\sigma^{\alpha} c_{\mathbf{q}-\mathbf{k}}, \quad \alpha=\pm,z,
\end{equation}
where $\varphi_\mathbf{q}$ is the pair wave function and reflects the similarity in construction of the jump operator and the corresponding pairing state. This gives rise to a Liouvillian in momentum space
\begin{equation}
\mathcal L [\rho] =\sum_{\alpha,\mathbf{k}} \left( 2J_{\mathbf{k}}^{\alpha}\rho J_{\mathbf{k}}^{\alpha\,\dag} - \{ J_{\mathbf{k}}^{\alpha\,\dag}J_{\mathbf{k}}^{\alpha}, \rho\}\right).
\end{equation}
The mean-field product ansatz for the density operator is
\begin{equation}
\rho = \prod_{\mathbf{q}}\rho_{ \mathbf{q}}, \quad \rho_{ \mathbf{q}} = \mathrm{tr}_{\neq  \mathbf{q}} \rho, \quad  \mathrm{tr} \rho_{ \mathbf{q}} =1\, \forall \,\,\mathbf{q} ,
\end{equation}
where each $\rho_{\mathbf{q}}$ spans the subspace for $\{\mathbf{q}\uparrow,-\mathbf{q}\downarrow\}$ \cite{Anderson58}. We thus allow for a residual entanglement of the momentum modes $\{\mathbf{q},-\mathbf{q}\}$, as in the BCS treatment, which is necessary to describe pairing. The second equation is the projection prescription to obtain the density operator for each mode which can be used to derive the equation of motion $\partial_t\rho_{ \mathbf{q}}$.

We take the partial trace $\mathrm{tr}_{\neq \mathbf{q}}$ on both sides of the master equation, for which we trace over all degrees of freedom outside the subspace $\{\mathbf{q}\uparrow,-\mathbf{q}\downarrow\}$. Then, in the spirit of BCS theory, we choose the relevant mean fields, i.e. macroscopically occupied expectation values in the dark state. As mean fields of linear and trilinear correlations vanish due to the Fermi statistics, and that of the quartic and higher order correlations connect momentum modes different from $\pm\mathbf{q}$, and thus are small compared to the macroscopic expectation values close to the steady state in the thermodynamic limit, we keep only mean fields of quadratic correlations, i.e. density mean fields or condensate with zero center of mass momentum [cf. Eqs. (\ref{MFExpVal}) (\ref{MFExpValgap1}) (\ref{MFExpValgap2})]. We thus see how the proximity to the steady state can be used as an ordering principle for a mean-field theory at late times, which is based on the exact knowledge of the fixed-phase steady state density matrix. Note that we use the commutation (as opposed to anti-commutation) properties during the process
\begin{eqnarray}\label{CommRules}
[\rho_{ \mathbf{p}},\rho_{\mathbf{ q}}] = [\rho_{\mathbf{ p}},c_{\mathbf{ q}}] = [\rho_{\mathbf{ p}},c_{ \mathbf{q}}^\dag]=0\quad {\rm for } \hspace{0.2cm} \mathbf{p}\neq \mathbf{q},
\end{eqnarray}
which is equivalent to the assumption that there is an even number of fermions in this mode.

We then have the following results:
\begin{eqnarray}\label{expression}
&&\mathrm{tr}_{\neq \mathbf{q}} \sum_{\mathbf{k}}\left( 2J_{\mathbf{k}}^{+}\rho J_{\mathbf{k}}^{+\,\dag} - \{ J_{\mathbf{k}}^{+\,\dag}J_{\mathbf{k}}^{+}, \rho\}\right) =\nonumber\\
&&2A (1 + |\varphi_{\mathbf{q}}|^2 ) \times \left\{ \gamma_{\mathbf{q},\downarrow} \rho_{\mathbf{q}}\gamma^{\dag}_{\mathbf{q},\downarrow} -\frac{1}{2}\{ \gamma^\dag_{\mathbf{q},\downarrow} \gamma_{\mathbf{q},\downarrow}, \rho_{\mathbf{q}}\}\right\},\nonumber\\
&&\mathrm{tr}_{\neq \mathbf{q}} \sum_{\mathbf{k}}\left( 2J_{\mathbf{k}}^-\rho J_{\mathbf{k}}^{-\,\dag} - \{ J_{\mathbf{k}}^{-\,\dag}J_{\mathbf{k}}^-, \rho\}\right) =\nonumber\\
&&2A (1 + |\varphi_{\mathbf{q}}|^2 ) \times \left\{ \gamma_{\mathbf{q},\uparrow} \rho_{\mathbf{q}}\gamma^\dag_{\mathbf{q},\uparrow} -\frac{ 1}{2}\{ \gamma^\dag_{\mathbf{q},\uparrow} \gamma_{\mathbf{q},\uparrow}, \rho_{\mathbf{q}}\}\right\},\nonumber\\
&&\mathrm{tr}_{\neq \mathbf{q}} \sum_{\mathbf{k}}\left( 2J_{\mathbf{k}}^z\rho J_{\mathbf{k}}^{z\,\dag} - \{ J_{\mathbf{k}}^{z\,\dag}J_{\mathbf{k}}^z, \rho\}\right) =
2A (1 + |\varphi_{\mathbf{q}}|^2 )\nonumber\\
&&\times \left\{ \gamma_{\mathbf{q},\uparrow} \rho_{\mathbf{q}}\gamma^\dag_{\mathbf{q},\uparrow}+ \gamma_{\mathbf{q},\downarrow} \rho_{\mathbf{q}}\gamma^\dag_{\mathbf{q},\downarrow} -\frac{1}{2}\{ \gamma^{\dag}_{\mathbf{q},\uparrow} \gamma_{\mathbf{q},\uparrow}+\gamma^\dag_{\mathbf{q},\downarrow} \gamma_{\mathbf{q},\downarrow}, \rho_{\mathbf{q}}\}\right\}.
\end{eqnarray}
where $A\equiv \int \frac{d\mathbf{q}}{(2\pi)^d} \frac{|\varphi_{\mathbf{q}}|^2}{1+|\varphi_{\mathbf{q}}|^2}\geq0$, and the integration runs over the first Brillouin zone.

Of particular interest and importance are the $\{\gamma_{\mathbf{q},\sigma}\}$ operators, which coincide with the definition of the Bogoliubov quasiparticles:
\begin{eqnarray}
\gamma_{\mathbf{q},\uparrow} &=& \frac{1}{\sqrt{1+|\varphi_\mathbf{q}|^2}} \, (c_{\mathbf{q},\uparrow} - \varphi_\mathbf{q} c^\dag_{-\mathbf{q},\downarrow})\nonumber ,\\
\gamma_{\mathbf{q},\downarrow} &=& \frac{1}{\sqrt{1+|\varphi_\mathbf{q}|^2}} \, (c_{-\mathbf{q},\downarrow} + \varphi_\mathbf{q} c^\dag_{\mathbf{q},\uparrow} ).\label{gamma}
\end{eqnarray}
Indeed, the mean -ield BCS pairing state is the vacuum state for the Bogoliubov
quasiparticles,
\begin{equation}
\gamma_{\mathbf{q},\uparrow} \vert D\rangle= \gamma_{-\mathbf{q},\uparrow}  \vert D\rangle= \gamma_{\mathbf{q},\downarrow} \vert D\rangle= \gamma_{-\mathbf{q},\downarrow}\vert D\rangle =0.
\end{equation}
Furthermore, when properly normalised as in Eq. (\ref{gamma}), these quasiparticle operators obey the closed fermion algebra:
\begin{eqnarray}
\{\gamma_{\mathbf{q},\sigma} , \gamma^\dag_{\mathbf{q}',\sigma'}\} &=& \delta_{\mathbf{q},\mathbf{q}'}\delta_{\sigma,\sigma'}, \nonumber\\
\{\gamma_{\mathbf{q},\sigma} , \gamma_{\mathbf{q}',\sigma'}\} &=& \{\gamma^\dag_{\mathbf{q},\sigma} , \gamma^\dag_{\mathbf{q}',\sigma'}\} =0,
\end{eqnarray}
i.e. they are related to the original fermion operators by a canonical transformation. We note that the operators $J^{a,z}_k$ do not exhibit a closed algebra structure.

From the expressions in Eq. (\ref{expression}), we may identify a damping spectrum in the dissipative part of the master equation:
\begin{eqnarray}
\kappa_\mathbf{q} =  2A\kappa  (1 + |\varphi_{\mathbf{q}}|^2 ),\label{dissgap}
\end{eqnarray}
with $A=(1-1/\sqrt{2})$ in 1-D, and $A\sim 0.36$ in 2-D. It is important to note that the damping spectrum is \emph{gapped}, i.e. $\kappa_\mathbf{q} \geq 2A\kappa$ is bounded from below. This behavior exhibits strong parallels to the equilibrium problem of paired fermions, where pairing is protected by an energy gap. Furthermore, it has the important implication that the approach to the d-wave dark state will be exponentially fast, in contrast to a bosonic system where wavelengths of arbitrary length make the approach to the dark states with long range order polynomially fast only \cite{Diehl08}. This result is reflected in the quantum trajectory simulations in the previous section.

Similarly, for the parent Hamiltonian of the pairing state, it is straightforward to derive
\begin{equation}
\mathrm{tr}_{\neq \mathbf{q}} [ H_p,\rho ]=[\sum_{\sigma}\kappa_{\mathbf{q}} \gamma^{\dag}_{\mathbf{q},\sigma}\gamma_{\mathbf{q},\sigma},\rho_{\mathbf{q}}]. \label{Heffdwave}
\end{equation}

Clearly, the dissipative mean-field theory developed here can be applied to pairing states with other spatial symmetries. As an important example, we discuss the result for complex p-wave pairing states. The p-wave pairing state for spinless fermions can be written as:
\begin{equation}
|P\rangle=\prod_{\mathbf{q}} (\frac{1}{\sqrt{1+|2\varphi_{\mathbf{q}}|^2}}+\frac{2\varphi_{\mathbf{q}}}{\sqrt{1+|2\varphi_{\mathbf{q}}|^2}}c^{\dag}_{\mathbf{q}}c^{\dag}_{-\mathbf{q}})|{\rm vac}\rangle,
\label{pwavestate}
\end{equation}
where $\mathbf{q}$ runs over half of first Brillouin zone, e.g. $q_x>0$. The pairing wavefunction $\varphi_{\mathbf{q}}^{\ast}=\varphi_{-\mathbf{q}}=-\varphi_{\mathbf{q}}$, which is required for p-wave symmetry: in 1-D, $\varphi_q=2i\sin q$; in 2-D, $\varphi_{\mathbf{q}}=2i(\sin q_x \pm i\sin q_y)$.

Following the previous derivations, we define the momentum space jump operators:
\begin{equation}
J_{\mathbf{k}}=\sum_{\mathbf{q}}\varphi_{\mathbf{q}}c^{\dag}_{\mathbf{q}}c_{\mathbf{q}-\mathbf{k}},
\end{equation}
where the summation over $\mathbf{q}$ is over the first Brillouin zone. Taking the partial trace over the degrees of freedom outside the subspace $\{\mathbf{q},-\mathbf{q}\}$ ($\mathbf{q}$ spans half the momentum space here, e.g. $q_x>0$), and identify the mean fields as before, we may arrive at the effective Hamiltonian
\begin{equation}
H_{\mathrm{eff}}=-\frac{i}{2}\sum_{\mathbf{q}}\kappa_{\mathbf{q}}\left(\gamma^{\dag}_{\mathbf{q}}\gamma_{\mathbf{q}}+\gamma^{\dag}_{-\mathbf{q}}\gamma_{-\mathbf{q}}\right),
\end{equation}
where the summation of $\mathbf{q}$ is over half the first Brillouin zone ($q_x>0$). The Bogoliubov quasi-particles are given as
\begin{equation}
\gamma_{\mathbf{q}}=\frac{1}{\sqrt{1+|2\varphi_{\mathbf{q}}|^2}}(c_{\mathbf{q}}-2\varphi_{\mathbf{q}}c^{\dag}_{-\mathbf{q}}).\label{pwavequasi}
\end{equation}
The dissipative coefficient is given as
\begin{equation}
\kappa_{\mathbf{q}}=A\kappa(1+|2\varphi_{\mathbf{q}}|^2),
\end{equation}
where  $A\equiv \int \frac{d\mathbf{q}}{(2\pi)^d} \frac{|\varphi_{\mathbf{q}}|^2}{1+|2\varphi_{\mathbf{q}}|^2}$, with the integral running over the first Brillouin zone. We find $A\sim 0.19$ in 1-D ($d=1$), and $A\sim 0.23$ in 2-D ($d=2$).

Finally, we note that the fermionic quasi-particle operators in Eq. (\ref{pwavequasi}) formally correspond to the Bogoliubov operators of a p-wave Hamiltonian in the limit $\mu\rightarrow-\infty$ ($\mu$ is the chemical potential), which means that the state is in the strong pairing limit  describing a state of delocalised tightly bound molecular pairs and is topologically trivial \cite{readgreen}. The generation of stable topological order is however possible in a modified setting, as has been established recently \cite{Diehl11}.

\section{Adiabatic passage to the ground state of the Hubbard model}\label{adHubbard}
\begin{figure}[tb]
\includegraphics[width=8cm]{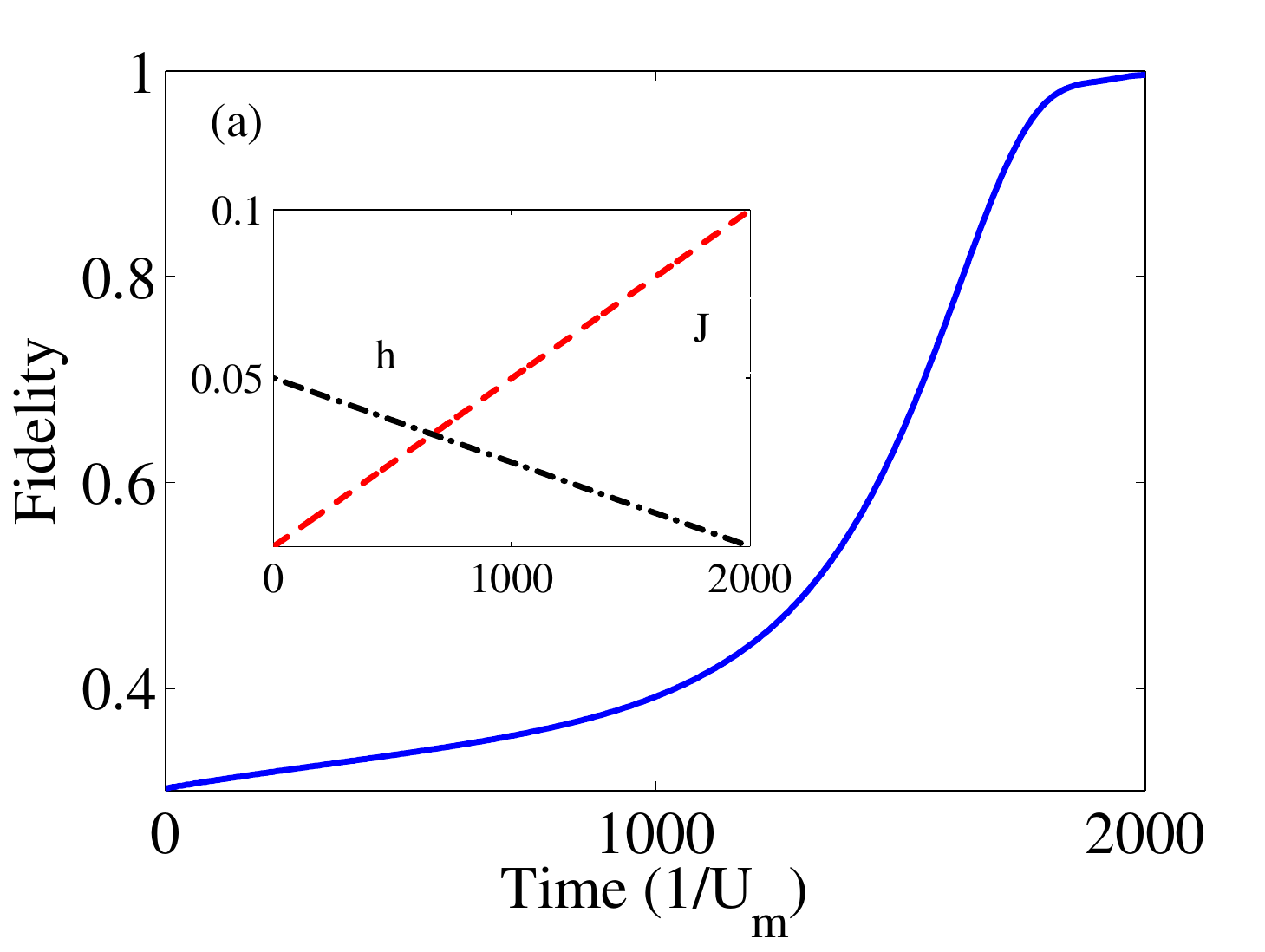}
\includegraphics[width=8cm]{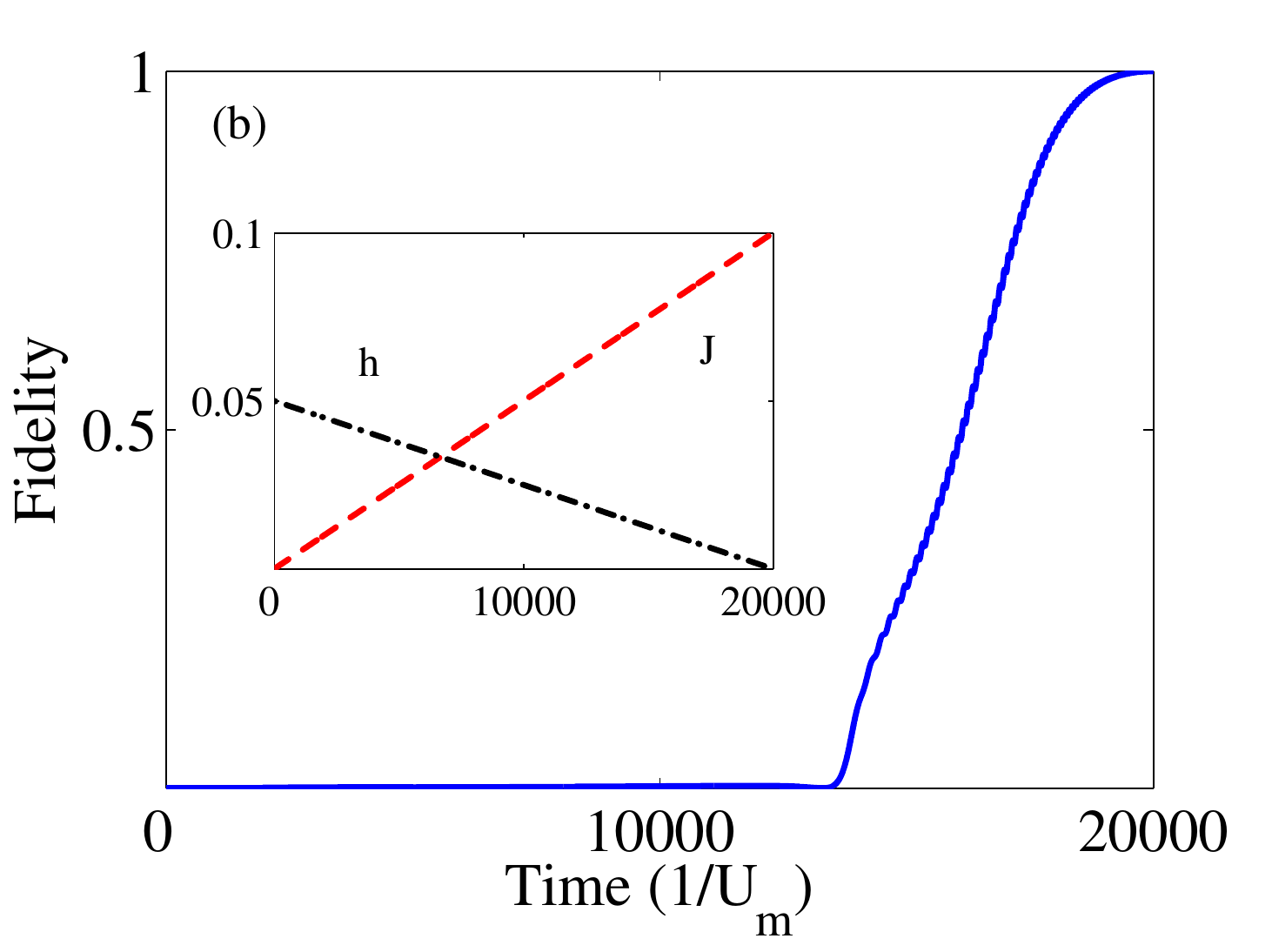}
\caption{Adiabatic passage connecting the antiferromagnetic state and the mean-field d-wave state with the ground state of the Fermi-Hubbard model. (a) The initial state is an antiferromagnetic N\'{e}el state on a $2\times 2$ plaquette with 4 atoms; (b) the initial state is a d-wave state on a $2\times 4$ ladder with $4$ atoms. We calculate the evolution of the fidelity of the instantaneous system state with respect to the final ground state of the FHM. (inset): Time dependence of the ramping parameters $h(t)/U_m$ and $J(t)/U_m$. The interaction energy $U(t)$ in the FHM linearly increases from $0$ to its maximum value $U_m$ during the ramp (not shown), with the final state corresponding to a strongly correlated situation with $J/U=0.1$. }
\label{Figadiabaticcomp}
\end{figure}
As argued above, the pure mean-field state with the correct symmetry (antiferromagnetic at half-filling and d-wave otherwise) is a convenient initial state for the quantum simulation of the ground state of the Fermi-Hubbard model (FHM). With a suitable adiabatic passage, it would be possible to connect the pure dark state with the ground state of the FHM. A first guess as to how to reach this ground state based on the experience with purely Hamiltonian systems might be a simple adiabatic passage, in which the Liouvillian is switched off while the FH Hamiltonian is ramped up. Small scale numerical simulations of the time evolution suggest that this procedure does not work for our combined system with both
unitary and dissipative evolution. This is understood from the fact that the mean-field state is not
an exact eigenstate of the FH Hamiltonian, such that unitary and dissipative evolutions compete. As a result of this competition, the steady state density matrix in general describes a mixed state, instead of a pure, zero entropy state \cite{Diehl08,Diehl10}. We thus observe that a dissipative gap cannot play the role of an energy gap in standard adiabatic passage schemes. We therefore propose a modified adiabatic passage, which uses the parent Hamiltonian of the mean-field state, constructed from the complete set of jump operators as in Eq. (\ref{parentH}). By construction, this parent Hamiltonian has the mean-field state as a gapped ground state, and therefore provides for an energetic stabilization. The passage now proceeds by first turning off the dissipation while the parent Hamiltonian is applied, then simultaneously ramping down the
parameters of the parent Hamiltonian while ramping up the parameters of the FHM. In this way, as long as the symmetry patterns of the mean-field target state and the ground state of the FHM Hamiltonian are the same and no phase transition is crossed, an energy gap persists through the whole passage. Indeed, we show numerically that this modified adiabatic passage ensures an efficient transfer into the desired ground state.

For the antiferromagnetic N\'{e}el state at half-filling, the parent Hamiltonian (dimensionless) reads
\begin{equation}
H_{\rm p}^{\rm AF }=\sum_{\langle i,j\rangle}(j_{i,j}^{\pm\,\dag}j_{i,j}^{\pm}+j^{\dag}_{i,j}j_{i,j})+j_{1}^{\dag}j_1,
\end{equation}
where the jump operators $j^{\pm}_{i,j}$ and $j_i$ are defined in Eqs. (\ref{1dafjump}, \ref{afjump2}). We have performed numerical simulations of such an adiabatic passage for a $2\times2$ plaquette with $4$ atoms. The result is shown in Fig. \ref{Figadiabaticcomp}(a). Indeed the initial N\'{e}el state can be adiabatically connected with the ground state of FHM at half-filling with high fidelity.

For the case of d-wave state, the parent Hamiltonian in Eq. (\ref{parentH}) by construction has the
initial d-wave state as an exact eigenstate and thus supports the d-wave state obtained from the
dissipative evolution. From Eq. (\ref{Heffdwave}), it is clear that the single fermion excitations on the d-wave state are gapped if the system is sufficiently far away from half-filling. As a consequence, all requirements for
an efficient adiabatic passage are met. Note that single fermion excitations above the ground state manifold of the reduced parent Hamiltonian are also suppressed by an energy gap, as $H_{\rm eff }^{\rm r }=\frac{1}{2}H_{\rm eff }$ on the mean-field level. We will make use of this important fact in Sec. \ref{altadiabatic} to design a modified adiabatic passage.

The time-dependent Hamiltonian describing the adiabatic passages is given as
\begin{equation}
H(t)=h(t)H_p+U(t)\sum_{i}c^{\dag}_{i\uparrow}c_{i\uparrow}c^{\dag}_{i\downarrow}c_{i\downarrow}-
J(t)\sum_{\langle i,j\rangle,\sigma}c^{\dag}_{i\sigma}c_{j\sigma},
\end{equation}
where the time dependent coefficients $h(t),U(t),J(t)$ give the precise path of the adiabatic passage. In practice, the time dependence of these coefficients is given by the rate at which the lattice potential giving rise to the FHM is ramped up, as well as by the rate of the effective interaction given by the parent Hamiltonian $H_p$. Here, for simplicity, we have chosen linear ramps for these coefficients (see insets of Fig. \ref{Figadiabaticcomp}), which already give a clear physical picture of the adiabatic passage. In practice these ramps could be further optimised, so that higher fidelities can be achieved in shorter ramp times. Consistent with the previous discussion, the role of the parent Hamiltonian is to provide an energy gap, and hence energetically stabilise the adiabatic passage.

We have performed numerical simulations of the adiabatic passage with various finite size systems. To avoid degeneracies of the ground state of FHM due to finite size effects, we have taken open boundary conditions for the Hubbard Hamiltonian during the adiabatic process, while we retain the periodic boundary conditions for the definition of the initial pairing state and for the jump operators. We expect that the mean-field d-wave state should be efficiently connected to the ground state of the Fermi-Hubbard Hamiltonian so long as the d-wave symmetry of the ground state of the Fermi-Hubbard model is present and not completely destroyed by the finite size effect. We find that this is the case for ladder systems. A typical result of our simulation on a finite ladder is shown in Fig. \ref{Figadiabaticcomp}(b). For systems in the thermodynamic limit, the symmetry property of the ground state is not affected by the boundary effect, and we expect an efficient adiabatic passage so long as the symmetries of the ground state are the same as the dissipatively driven initial state \cite{symmetryrev,Shen,correlation}.

\section{Physical implementation and modified adiabatic passage}

As an illustrative example, we now discuss a proof-of-principle implementation of the single-particle jump operators. The scheme we describe in this section is stroboscopic, and involves realising the action of the jump operators in a series of steps. Though non-trivial to implement in present experiments, this example is made up of elements that are presently accessible in experiments.  The example illustrates how the properties of the operators appearing in the previous sections, specifically that they are quasi-local, conserve particle number, and can be implemented based on single-particle operations, make them favourable for experimental implementation. For an alternative non-stroboscopic, i.e. ``always-on'' continuous implementation, which is applicable in the case of spinless (spin-polarised) fermions such as the p-wave case discussed above, see \cite{Diehl11}.

\begin{figure}[tb]
\centerline{\includegraphics[width=8cm]{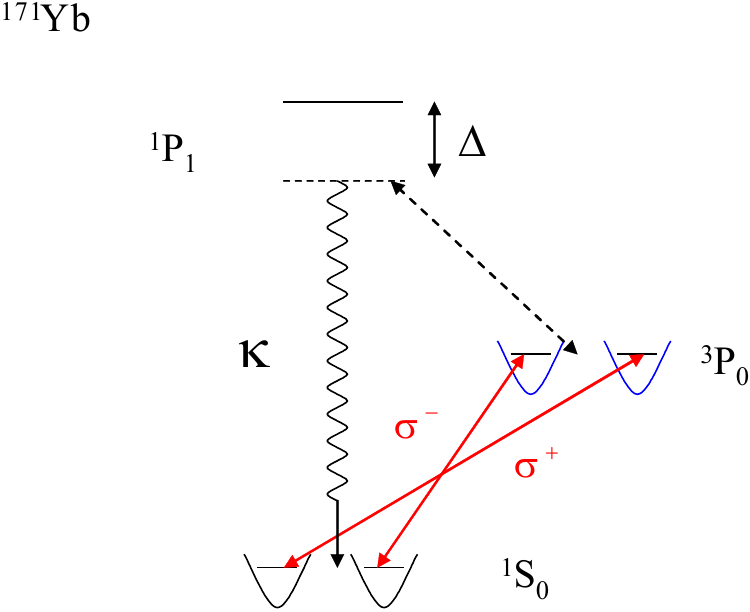}}
\caption{Level scheme using $^{171}$Yb atoms.
The physical spin state is encoded in the nuclear spin sublevels of the $^1$S$_0$ manifold. The spin flip operation is implemented via
off-resonant coherent coupling to the $^3$P$_0$ manifold with circularly polarised light (red arrows).
The long lived $^3$P$_0$ states are coupled to the $^1$P$_1$ level in a two-photon process,
from which spontaneous emission into a cavity is induced, leading back to the $^1$S$_0$ manifold.}
\label{Figlevel}
\end{figure}

Our example takes advantage of the properties of alkaline-earth-like atoms \cite{AEatom1,AEatom2,AEatom3,Deutsch,Gorshkov,aereview}. With two valence electrons, these atoms possess metastable triplet levels, and fermionic isotopes have non-zero nuclear spin (e.g., $I=1/2$ for $^{171}$Yb, which we will choose here). This nuclear spin acts as an independent degree of freedom in the ground $^1$S$_0$ and lowest excited $^3$P$_0$ manifolds. Here, the nuclear spin will play the role of the physical fermionic spin degree of freedom, and the $^3$P$_0$ manifold will be used as an intermediate state in the dissipative process. These states are depicted in Fig. \ref{Figlevel}. Note that as  $^1$S$_0$ and $^3$P$_0$ are optically separated, they can be trapped in independent lattices using dipole traps at different wavelengths \cite{AEgate}.

As a simple example, we will first discuss the implementation of jump operators for driving the system into the antiferromagnetic N\'{e}el state at half-filling. We will then move on to the more complicated cases of pairing states.

\subsection{Antiferromagnetic N\'{e}el state}

\begin{figure}[tb]
\centerline{\includegraphics[width=10cm]{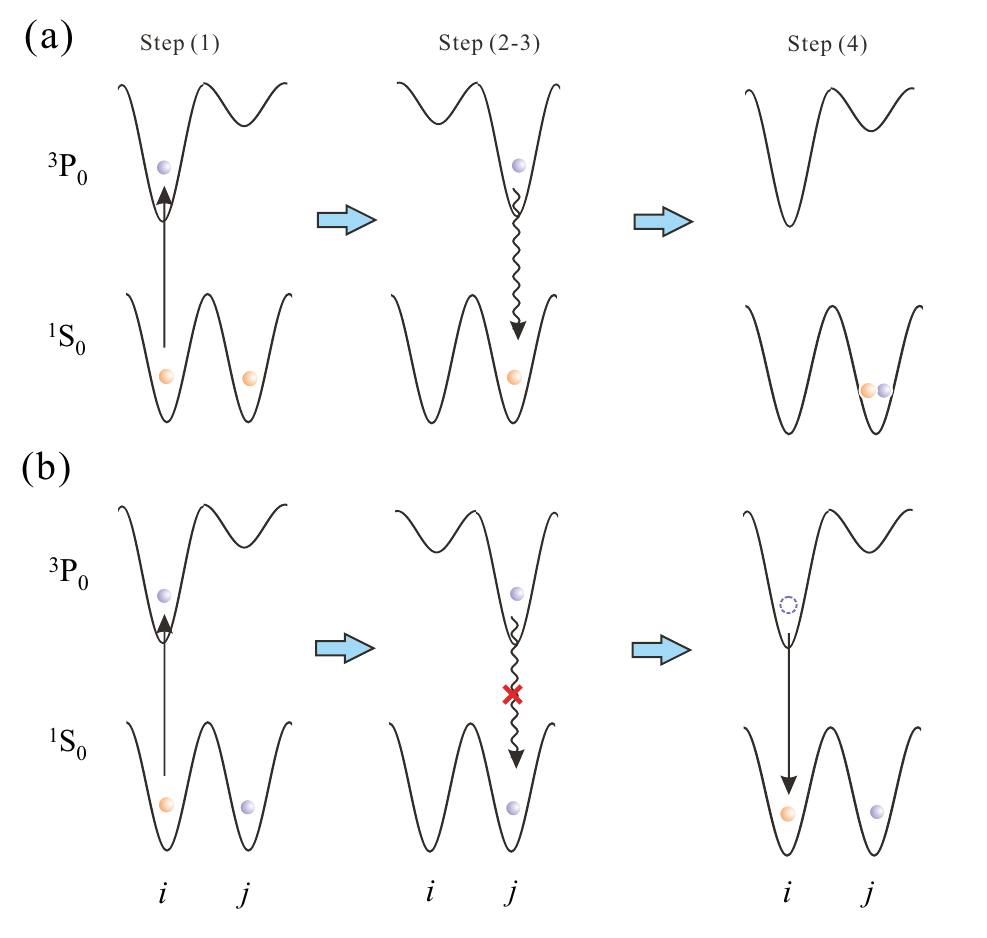}}
\caption{Implementation scheme for the jump operator $c_{j,\uparrow}^\dag c_{i,\downarrow}$ for 1-D antiferromagnetic N\'{e}el state. (a) The decay channel given by the jump operator is not blocked; (b) the decay channel is Pauli blocked. }
\label{Figafimplement}
\end{figure}

Since the Lindblad jump operators for the N\'{e}el states act on unit cells of two sites, we will focus on operations on two adjacent sites $i$ and $j$. These will be carried out in parallel on pairs of adjacent sites along the lattice. As illustrated in Fig. \ref{Figlevel}, spins states are encoded in the nuclear spin sublevels of the $^1$S$_0$ manifold. We further assume that atoms in the $^1$S$_0$ and $^3$P$_0$ manifolds are trapped in independent optical lattices, and that $^3$P$_0$ is trapped in a superlattice with a period of two sites, defining pairs of sites, where we label the left well $i$ and the right well $j$. Initially, the superlattice potential is arranged in such a way that the potential well at site $i$ is much deeper than at site $j$ (with an energy difference of the lowest state in each well of the order of several kHz). The action of the jump operator $c^{\dag}_{j,\uparrow}c_{i,\downarrow}$ can then be realised by performing the following operations (see Fig. \ref{Figafimplement}): $(1)$ apply a circularly polarised $\pi$-pulse selectively on site $i$ coupling the $^1$S$_0$ and $^3$P$_0$ manifolds so that any atom originally in the state $|\!\downarrow,^1{\rm S}_0\rangle$ will end up in $|\!\uparrow, ^3{\rm P}_0\rangle$; $(2)$ adiabatically manipulate the superlattice potential so that the population at site $i$ is transferred to site $j$; $(3)$ couple the $^3$P$_0$ and $^1$P$_1$ manifolds off-resonantly, so that the population in $|\!\uparrow, ^3{\rm P}_0\rangle$ should decay to $|\!\uparrow,^1{\rm S}_0\rangle$ on site $j$ if and only if the state on site $j$ is empty; $(4)$ repeat the steps $(2)$ and then $(1)$ (in reverse order) to bring any remaining population in $^3$P$_0$ back to the ground state manifold.

Before moving on to extend the scheme to jump operators associated with pairing states, several comments are in order: (a) the jump operator is implemented stroboscopically, which places requirements on the time scale of each step of operations listed above, such that the total time of evolution should be much longer than the time scale of operations; (b) the jump operators are implemented in parallel for each pair of lattice sites along the lattice; (c) during the excitation of the population from $^1$S$_0$ to $^3$P$_0$, as the line-width of the metastable $^3$P$_0$ is on the order of $10$mHz for $^{171}$Yb, the bias between different subwells in the superlattice potential ensures site selectivity; (d) the nuclear spin is conserved during the decay process, which is guaranteed by the large detuning from the $^1$P$_1$ manifold. The nuclear spin conservation can also be realised in this case by applying a large magnetic field so that electronic spin and nuclear spin are decoupled \cite{Deutsch}.

\begin{figure}[tb]
\centerline{\includegraphics[width=10cm]{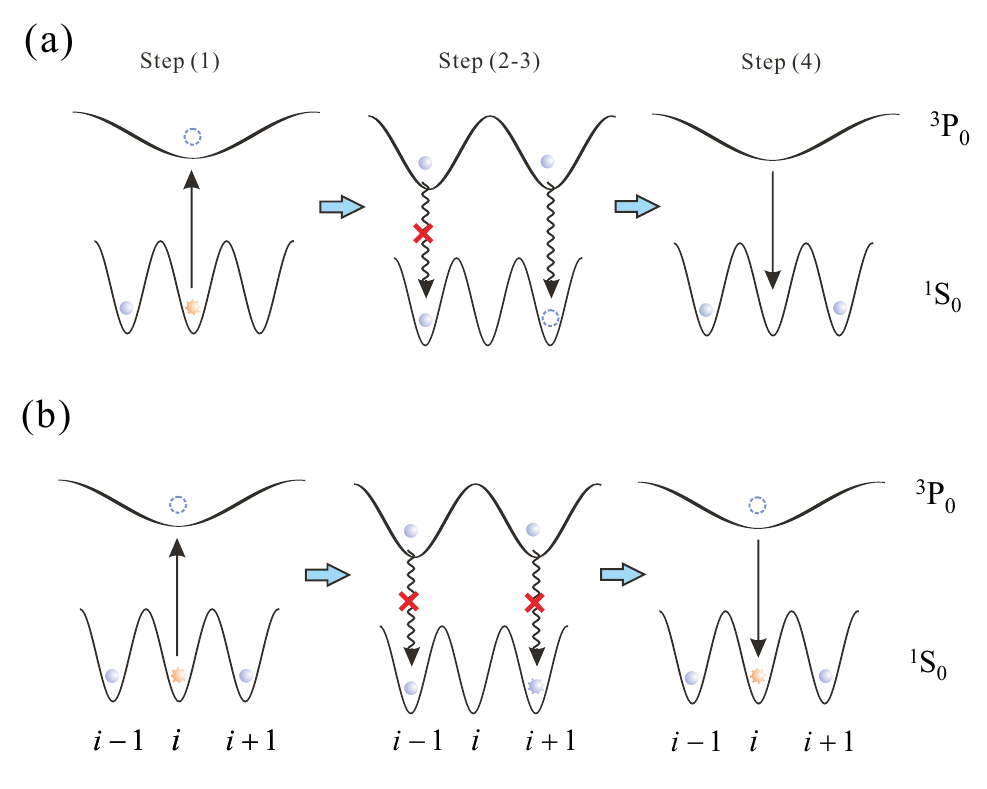}}
\caption{Implementation scheme for jump operators of 1-D d-wave state. (a) Only one of the decay channels is blocked; (b) both decay channels are Pauli blocked, therefore the jump operator does not change the system configuration. The state is a local dark state for this local jump operator.}
\label{Fig1dwaveimplement}
\end{figure}

\subsection{d-wave pairing state}

We now extend the ideas above to the implementation of jump operators for driving the system into d-wave pairing states. For the d-wave jump operators, an additional constraint on the dissipative process is that atoms on quasi-local sites, e.g. site $i+e_x$ and $i-e_x$, should decay coherently. To satisfy this requirement, we couple the system to a cavity with a finite linewidth. An atom (or atoms) at the sites $i+e_x$ and $i-e_x$ will then be coupled collectively to the cavity mode, ensuring that the decay is coherent. For clarity, we first describe the implementation procedures in 1-D, and choose the example of $J_{i}^{+} = (c_{i+ 1,\uparrow}^\dag + c_{i- 1,\uparrow}^\dag )c_{i,\downarrow}$.  The step-by-step implementation scheme is shown in Fig. \ref{Fig1dwaveimplement}: $(1)$ We first assume that the $^3$P$_0$ state is initially trapped in a lattice of three times the period as that for the $^1$S$_0$ state, defining blocks of three sites in the original lattice. Using this, we excite any spin-down atom in $^1$S$_0$ on central site to the spin-up state of the $^3$P$_0$ manifold, using $\sigma^+$ light. $(2)$ We then add an additional potential, splitting this site into two, and separate these sites so that the mode of atoms confined in them overlap the right and left sites of the original three-site block. $(3)$ We induce decay by coupling atoms in the $^3$P$_0$ state off-resonantly to the $^1$P$_1$ state, with coupling strength $\Omega$, and detuning $\Delta$. If we couple the $^1$S$_0$--$^1$P$_1$ transition to a cavity mode with linewidth $\Gamma$ and vacuum Rabi frequency $g$, then the decay will be coherent over the triple of sites. In the limit $\Delta\gg \Omega, \Gamma$ and $\Gamma\gg \max( \frac{\Omega^2}{\Delta},\frac{g^2}{\Delta},\frac{\Omega g}{\Delta})$, we adiabatically eliminate the cavity mode and the intermediate far off-resonant state $^1$P$_1$, and obtain an effective decay rate $\Gamma_{\rm eff}=\frac{\Omega^2g^2}{\Delta^2\Gamma}\sim 9$kHz for typical parameters (see \ref{appendixc}). Note that Fermi statistics will be observed in this process, and that we assume that the atoms remain in the lowest band, as all parameters are smaller than the trapping frequency in the lattice (see Ref.~\cite{sandner2011} for more details of Pauli-blocking of spontaneous emissions in this sense).

Other jump operators, $J_{i}^-$ and $J_{i}^z$ can be implemented by applying rotations in the nuclear spin before and after the three steps above. For $J_{i}^-$, one exchanges the spins with a $\pi$-pulse, whereas for $J_{i}^z$, one must apply a $\pi/4$ rotation in the nuclear spin basis before and after the operation. In addition, for $J^z_{i}$, both spin states should be excited, and coherence of nuclear spins is maintained throughout the operation. This can be achieved by either going far off-resonant for the field coupling $^3$P$_0$ and $^1$P$_1$ manifold, or by applying a large magnetic field as described in Ref. \cite{Deutsch}.

\begin{figure}[tb]
\centerline{\includegraphics[width=10cm]{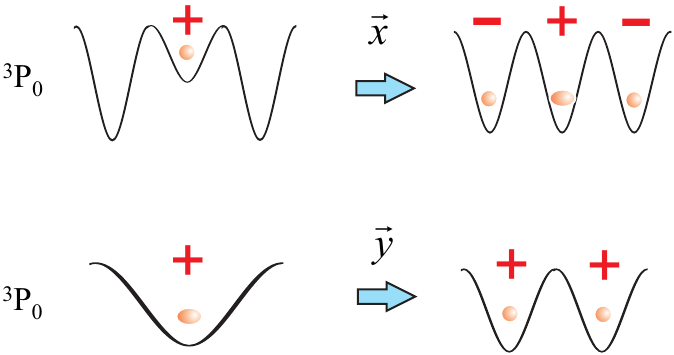}}
\caption{Generalisation of the implementation of d-wave jump operators to a 2-D lattice. Only the manipulation of the upper superlattice is shown here. (upper panel) After the population in the central site of the 3-by-3 plaquette in the $^1$S$_0$ level is excited to the superlattice of $^3$P$_0$, the potential at the central site is adiabatically lowered so that the state is adiabatically connected to the one where the relative phase between the central site and its neighbors is negative; (lower panel) the superlattice is then shifted adiabatically in the $\mathbf{y}$ direction, splitting the remaining population in the central site along $\mathbf{y}$ so that the correct relative phase with d-wave symmetry is imposed as given by $\rho_\nu$ in Eq. (\ref{dunitcell}). One may then follow the procedure for 1-D singlet pairing state implementation.}
\label{Fig2dwaveimplement}
\end{figure}

This scheme can be generalised to 2-D by considering 3-by-3 plaquettes defined by the appropriate superlattice potential for the $^3$P$_0$ level. As in the 1-D case, we require an adiabatic manipulation of this potential in step (ii), although here the depths of the wells must be adjusted to ensure that the correct relative phases are obtained for atoms ``transported'' in different directions (see Fig. \ref{Fig2dwaveimplement} and its caption).

\subsection{Implementing the reduced parent Hamiltonian and modified adiabatic passage}\label{altadiabatic}

Here, we extend the scheme above to implement the \emph{reduced} parent Hamiltonian for the d-wave state stroboscopically. We see from the discussion in Sec. \ref{symmetryanalysis} that the mean-field d-wave state is in the ground state manifold of the reduced parent Hamiltonian.  As we have discussed in Sec. \ref{adHubbard}, this degenerate ground state manifold (two-fold in 1-D, four-fold in 2-D) is protected by an energy gap from single fermion excitations under the reduced parent Hamiltonian. Furthermore, we will also show below that an adiabatic passage with high fidelity can be achieved by a modified adiabatic passage scheme.

\begin{figure}[tb]
\centerline{\includegraphics[width=10cm]{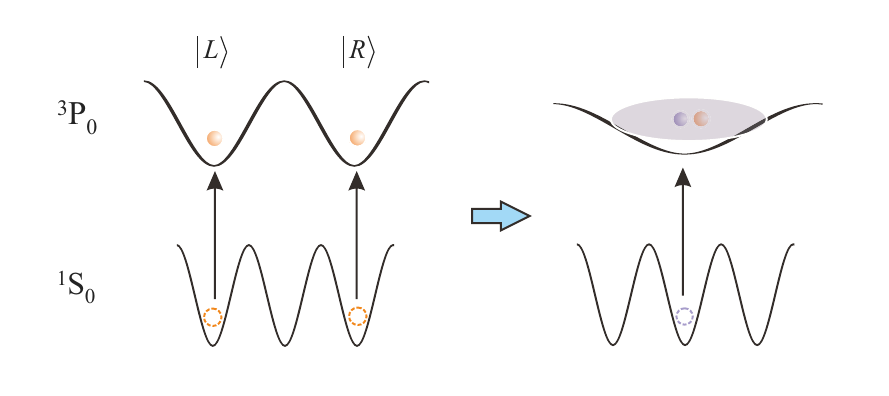}}
\caption{Implementation of $(J_{i}^{\pm})^{\dag}J_{i}^{\pm}$ in the parent Hamiltonian for 1-D case. Firstly, the population of the outer sites are excited to the upper lattice. The super lattice is then adiabatically tuned from a double well structure to a single well, so that the state $|L\rangle+|R\rangle$ is projected to the lowest level of the single well potential. The interaction is then induced via a Feshbach resonance for instance after exciting the population of the opposite spin in the central site to the superlattice.}
\label{FigparentHimplement}
\end{figure}

The implementation of the reduced parent Hamiltonian is similar to that of the jump operators, except
that the dissipative part is replaced by an induced phase shift. As shown in Eq. (\ref{redparentH}), the reduced parent Hamiltonian contains an effective interaction term and a term proportional to the total particle number that is not important for states with fixed particle number. To implement the effective interaction term stroboscopically, as illustrated in Fig. \ref{FigparentHimplement} with the example of 1-D Hamiltonian $(J_i^{+})^{\dag}J_{i}^{+}$, the following steps are required: $(1)$ any spin down atoms in the left and right well
of the ground state lattice potential are transferred to the superlattice potential of the $^3$P$_0$ manifold; $(2)$ the
double-well in the superlattice potential is merged into a single well, during which process the symmetric state in
the double-well potential is mapped to the lowest motional state of the final single well
potential; $(3)$ a phase shift is then induced to generate an interaction only if the spin-down state in
the $^1$S$_0$ manifold and the spin-up state in the $^3$P$_0$ manifold are simultaneously occupied. This
can be achieved, e.g., by applying a $\pi$ pulse between the spin-down state in the $^1$S$_0$
state and the spin-down state in the lowest motional state in the superlattice potential, and
inducing an interaction between the different spin states in the $^3$P$_0$ manifold via an optical
Feshbach resonance. Finally, to implement $(J^{-}_i)^{\dag}J_{i}^{-}$, the spins should be exchanged while the above procedure is carried out.

\begin{figure}[tb]
\includegraphics[width=8cm]{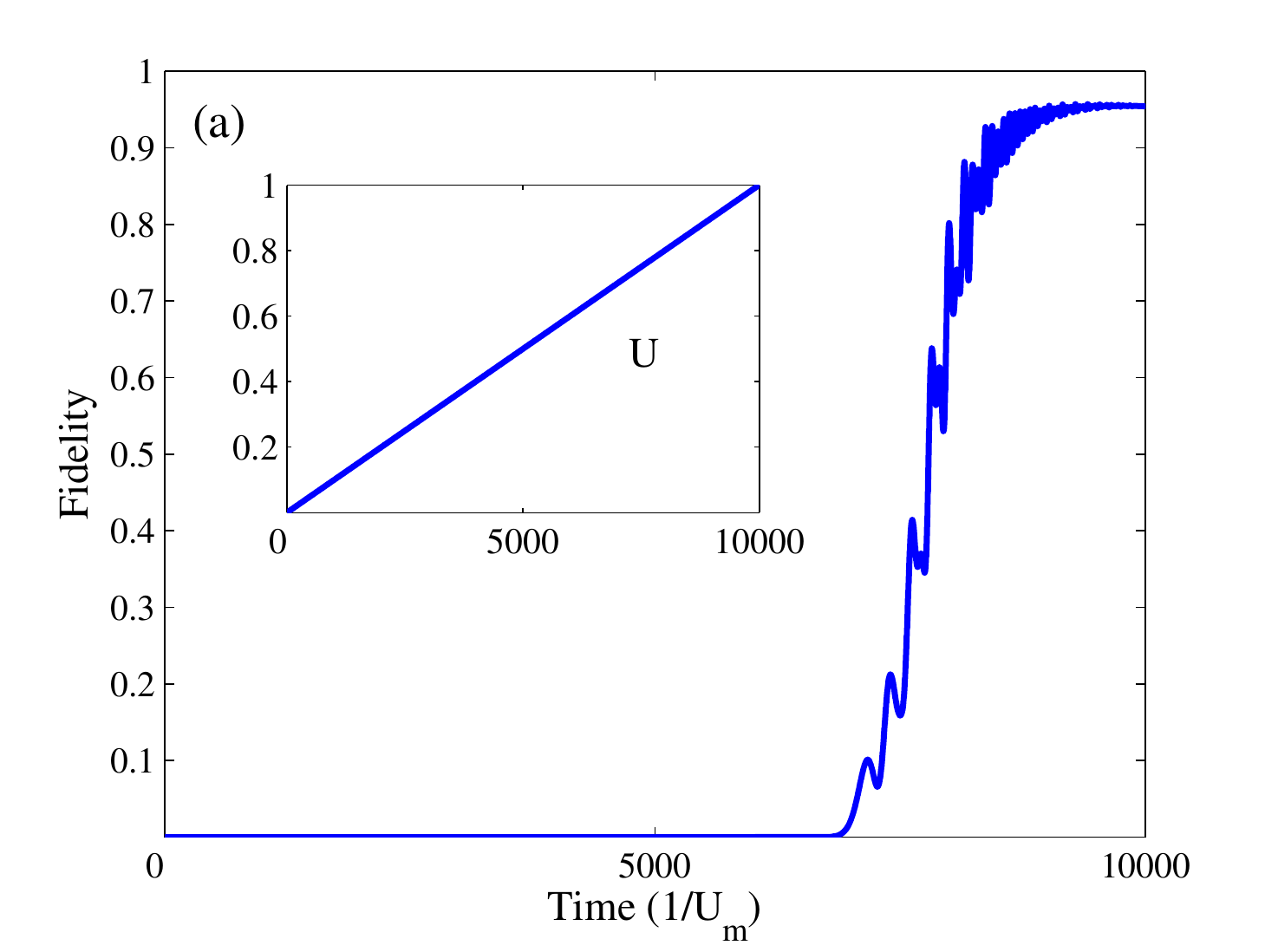}
\includegraphics[width=8cm]{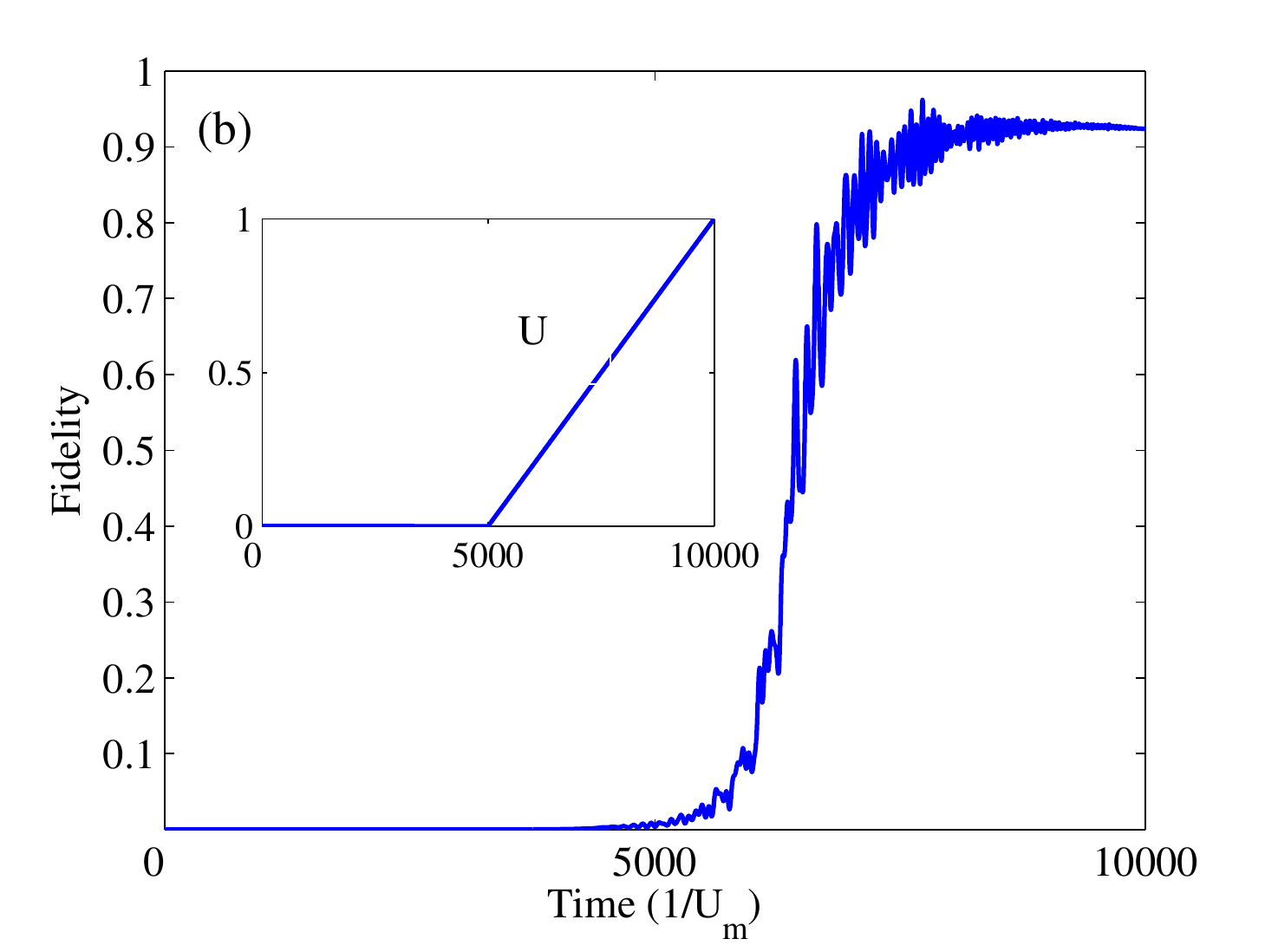}
\caption{Adiabatic passage connecting the mean-field d-wave state with the ground state of the Fermi-Hubbard model on a $2\times6$
ladder with 4 atoms with (a) complete parent Hamiltonian; (b) reduced parent Hamiltonian.
We calculate the evolution of fidelity of the system state with respect to the final ground state of the FHM. (inset): Time dependence of the interaction rate $U(t)/U_m$. In both cases, $h(t)/U_m$ is ramped down linearly from $0.05$ to $0$, while $J(t)/U_m$ is ramped up from $0$ to $0.1$. Similar to Fig. \ref{Figadiabaticcomp}(b), the fidelity remains small until late in the adiabatic process. The overlap with the ground state of the FHM only becomes large at late times when the repulsive interaction is large enough to overcome double-occupancies that occur due to the form of
the mean-field pairing wavefunction.}
\label{Figadiabatic264}
\end{figure}

With only the reduced parent Hamiltonian, we find that given an optimised ramping scheme, the mean-field d-wave state can still be adiabatically connected with the ground state of FHM on small lattices. Fig. \ref{Figadiabatic264} shows such an example, where ramps with the complete parent Hamiltonian and with the reduced parent Hamiltonian are numerically simulated for 4 atoms on a 2$\times$6 ladder. For the adiabatic passage with the reduced parent Hamiltonian, we ramp up $J(t)$ and $U(t)$ separately. In both cases, we have very high fidelity at the end of the ramp. For ramps with the reduced parent Hamiltonian, the high fidelity is due to the large overlap ($\sim 0.95$ for most ladder systems) between the mean-field d-wave state and the ground state of the time dependent Hamiltonian at the beginning of the ramping process (when there is no on-site interaction). For the numerical simulations that we considered here, this overlap also sets the upper bound for the final fidelity of the ramps.

\section{Conclusions}

We have proposed an approach for the preparation of many-body pairing states of given symmetry for fermionic atoms in an optical lattice via driven dissipative processes based on suitable reservoir engineering. We have discussed in detail the strategy of designing the jump operators making use of the Fermi statistics, which gives rise to the dissipative preparation of the initial pairing state. This process is in general efficient, due to the existence of a dissipative gap. We then argued for the uniqueness of the pairing state as the steady state of the dissipative dynamics, both from symmetry considerations, and via small scale numerical simulations. Note that for realistic finite size systems such as plaquette geometries \cite{finitesizeETH}, it is also possible to design jump operators for specific many-body states defined on the finite system, in which case one may need to design special ``boundary" jump operators to make the state unique.

We then discussed the adiabatic passage process that could be used to connect the driven-dissipative mean-field state with the ground state of the FHM. As our d-wave state is not an eigenstate of the FHM, directly ramping down the dissipation rate while ramping up the FHM leads to competition between the coherent and dissipative dynamics which would not drive the system into the ground state. We therefore introduced the parent Hamiltonian of the d-wave state, a semi-positive Hermitian Hamiltonian constructed from the jump operators. By construction, the parent Hamiltonian has the dark state of the dissipative process as its ground state. We illustrated via small scale numerical simulations that the ground state of the FHM can be adiabatically connected with the mean-field state of the relevant symmetry via optimised adiabatic paths. This is in similar spirit to the recent experimental demonstration of antiferromagnetic order in an optical lattice \cite{Greinerqs}, where the desired eigenstate of the Ising model is prepared via adiabatic passage from a starting state that has low entropy and sufficient overlap with the final state. We note that it is possible to extend these small scale numerical calculations by applying time dependent density matrix renormalisation group (t-DMRG) methods. In fact, quantum trajectories methods could be combined with t-DMRG methods \cite{Dal09} in order to perform larger-scale simulations of the dissipative preparation process and the adiabatic ramp together.

Finally, we discussed a proof-of-principle physical implementation of both the jump operators and the parent Hamiltonian using alkaline-earth-like atoms, which illustrated that the properties of the jump operators discussed here are favourable for implementation. We mainly focused on the implementation of d-wave pairing state, but similar implementations can be readily found for pairing states of other symmetries, so long as the jump operators are quasi-local and involve operations manipulating only single particles.

\ack
We thank A. Gorshkov, K. Hammerer, B. Kraus, A. Kantian and M. Foss-Feig for discussions. This work was supported by the Austrian Science Fund through SFB F40 FOQUS and EUROQUAM\_DQS (I118-N16), the EU through IP AQUTE, NSFC (11105134), and The Fundamental Research Funds for the Central Universities (WK2470000001,WK2470000006).

\section*{Appendix}
\appendix
\section{Two-particle jump operators for d-wave pairing state}\label{appendnew}

In this appendix, we derive in detail the two-particle jump operators for pairing states with d-wave symmetry as appeared in Eqs. (\ref{jump},\ref{twoparticle2d}).

\subsection{One-dimensional case}\label{1djumpdesign}

As an example, we choose
\begin{equation}
\xi_i=c_{i+1\downarrow}c_{i\uparrow}-c_{i\downarrow}c_{i+1\uparrow}.
\end{equation}
The commutation relations then give
\begin{eqnarray}
A_i&=&\left\{(c^{\dag}_{i+1\uparrow}c_{i+1\uparrow}-c^{\dag}_{i+1\downarrow}c_{i+1\downarrow})
-(c^{\dag}_{i\uparrow}c_{i\uparrow}-c^{\dag}_{i\downarrow}c_{i\downarrow})\right.\nonumber\\
&+&\left.(c^{\dag}_{i-1\uparrow}c_{i+1\uparrow}-c^{\dag}_{i-1\downarrow}c_{i+1\downarrow})
-(c^{\dag}_{i+2\uparrow}c_{i\uparrow}-c^{\dag}_{i+2\downarrow}c_{i\downarrow})\right\}\\
B_i&=&2\left\{(c^{\dag}_{i+1\uparrow}c^{\dag}_{i+2\downarrow}-c^{\dag}_{i+2\uparrow}c^{\dag}_{i+1\downarrow})
+(c^{\dag}_{i+1\uparrow}c^{\dag}_{i\downarrow}-c^{\dag}_{i\uparrow}c^{\dag}_{i+1\downarrow})\right.\nonumber\\
&+&\left.(c^{\dag}_{i-1\uparrow}c^{\dag}_{i\downarrow}-c^{\dag}_{i\uparrow}c^{\dag}_{i-1\downarrow})
+(c^{\dag}_{i-1\uparrow}c^{\dag}_{i+2\downarrow}-c^{\dag}_{i+2\uparrow}c^{\dag}_{i-1\downarrow})\right\}.
\end{eqnarray}
It is straightforward to show that if we define
\begin{eqnarray}
\chi_i&=&\left\{(c^{\dag}_{i+1\uparrow}c^{\dag}_{i+2\downarrow}+c^{\dag}_{i+2\uparrow}c^{\dag}_{i+1\downarrow})
+(c^{\dag}_{i+1\uparrow}c^{\dag}_{i\downarrow}+c^{\dag}_{i\uparrow}c^{\dag}_{i+1\downarrow})\right.\nonumber\\
&+&\left.(c^{\dag}_{i-1\uparrow}c^{\dag}_{i\downarrow}+c^{\dag}_{i\uparrow}c^{\dag}_{i-1\downarrow})
+(c^{\dag}_{i-1\uparrow}c^{\dag}_{i+2\downarrow}+c^{\dag}_{i+2\uparrow}c^{\dag}_{i-1\downarrow})\right\},
\end{eqnarray}
then $\chi_iB_i=0$.

The symmetry in the expressions above suggest that we may simplify these operators
by choosing $\xi_i=c_{i+1\downarrow}c_{i\uparrow}$. The commutation relations then have the form:
\begin{eqnarray}
A_i&=&1-c_{i\uparrow}^{\dag}c_{i\uparrow}-c^{\dag}_{i+1\downarrow}c_{i+1\downarrow}
-c^{\dag}_{i+2\uparrow}c_{i\uparrow}
-c^{\dag}_{i-1\downarrow}c_{i+1\downarrow}\label{1dAi}\\
B_i&=&-2(c^{\dag}_{i\uparrow}+c^{\dag}_{i+2\uparrow})(c^{\dag}_{i+1\downarrow}+c^{\dag}_{i-1\downarrow}).\label{1dBi}
\end{eqnarray}
The most straightforward choice of $\chi_i$ would be $\chi_i=B_i$, as $B_i^2=0$. More
generally, $\chi_iB_i=0$ is satisfied so long as the pair operators in $\chi_i$ can be factored
out to contain either $(c^{\dag}_{i\uparrow}+c^{\dag}_{i+2\uparrow})$ or
$(c^{\dag}_{i+1\downarrow}+c^{\dag}_{i-1\downarrow})$. This actually allows some freedom in
choosing the remaining part of the $\chi_i$ operator.

However, in this second scenario, the existence of a constant term in Eq.
(\ref{1dAi}) renders Eq. (\ref{gencomm}) not equal to zero even if Eq. (\ref{cond}) is satisfied.
The resulting jump operator would then not give the desired dark state. To solve this problem, one
needs to introduce appropriate symmetry into the design of the jump operator. Notice that assuming
the translational symmetry, the creation operator of the state can also be written as:
\begin{equation}
\eta_j=c^{\dag}_{j\uparrow}c^{\dag}_{j+1\downarrow}+c^{\dag}_{j\uparrow}c^{\dag}_{j-1\downarrow}
=c^{\dag}_{j\uparrow}(c^{\dag}_{j+1\downarrow}+c^{\dag}_{j-1\downarrow}).
\end{equation}
Correspondingly, we examine the following factorised $\xi_i$:
\begin{equation}
\xi_i=(c_{i+1\downarrow}-c_{i-1\downarrow})c_{i\uparrow}.
\end{equation}
Note that the choice of the negative sign here is to ensure that no constant terms appear in the
expression for $A_i$.

For the commutation relations, we now have
\begin{eqnarray}
A_i&=&(c^{\dag}_{i+1\downarrow}c_{i-1\downarrow}-c^{\dag}_{i-1\downarrow}c_{i+1\downarrow})\nonumber\\&+&
(c^{\dag}_{i-1\downarrow}c_{i-1\downarrow}-c^{\dag}_{i+1\downarrow}c_{i+1\downarrow})
+(c^{\dag}_{i-2\uparrow}c_{i\uparrow}-c^{\dag}_{i+2\uparrow}c_{i\uparrow})\\
B_i&=&2(c^{\dag}_{i-2\uparrow}-c^{\dag}_{i+2\uparrow})(c^{\dag}_{i-1\downarrow}+c^{\dag}_{i+1\downarrow}).
\end{eqnarray}
This implies that we may satisfy the dark state requirement by choosing a jump operator of the form
\begin{equation}
J_i=C^\dag  M c_{i\uparrow}, \quad M = (c^{\dag}_{i+1\downarrow}+c^{\dag}_{i-
1\downarrow})(c_{i+1\downarrow}-c_{i-1\downarrow}),
\end{equation}
as given in Eq. (\ref{jump}).

\subsection{Two-dimensional case}

We define
\begin{equation}
\xi_i= (c_{i+e_x\downarrow} +c_{i+e_y\downarrow}) c_{i\uparrow},\label{2djump}
\end{equation}
whose commutation relations are:
\begin{eqnarray}
A_i&=&\left\{\left[(c^{\dag}_{i+e_y\downarrow}+c^{\dag}_{i-e_y\downarrow})c_{i+e_y\downarrow}-(c^{\dag}_{i+e_x\downarrow}+c^{\dag}_{i-e_x\downarrow})c_{i+e_x\downarrow}\right]\right.\nonumber\\
&\times&\left[(c_{i+e_y\downarrow}^\dag+c_{i-e_y\downarrow}^\dag) c_{i+e_x\downarrow}-(c^{\dag}_{i+e_x\downarrow}+c^{\dag}_{i-e_x\downarrow})c_{i+e_y\downarrow}\right]\nonumber\\
&+&\left.(c^{\dag}_{i+2e_y\uparrow}c_{i\uparrow}-c^{\dag}_{i+2e_x\uparrow}c_{i\uparrow})
+(c^{\dag}_{i+e_x-e_y\uparrow}c_{i\uparrow}-c^{\dag}_{i+e_y-e_x\uparrow}c_{i\uparrow})  \right\}\\
B_i&=&2(c^{\dag}_{i+e_y-e_x\uparrow}-c^{\dag}_{i+e_x-e_y\uparrow}-c^{\dag}_{i+2e_y\uparrow}+c^{\dag}_{i+2e_x\uparrow})\nonumber\\
&\times&(c^{\dag}_{i+e_y\downarrow}+c^{\dag}_{i-e_y\downarrow}-c^{\dag}_{i+e_x\downarrow}-c^{\dag}_{i-e_x\downarrow}).
\end{eqnarray}

Following the same derivation as in the previous section, we find
\begin{equation}
\chi_i=C^\dag(c^{\dag}_{i+e_y\downarrow}+c^{\dag}_{i-e_y\downarrow}-c^{\dag}_{i+e_x\downarrow}-c^{\dag}_{i-e_x\downarrow}),
\end{equation}
where $C^\dag$ is an arbitrary superposition of single-fermion creation operators. Note that this is the
most straightforward choice to satisfy $\chi_iB_i=0$, other solutions may still exist.

Finally, we see that in the case of a d-wave state on a 2-D lattice, the jump
operator takes the form:
\begin{eqnarray}
J_i&=&C^\dag  M c_{i\uparrow}\nonumber\\
M &=& -\sum_{\nu}\rho_{\nu}c^{\dag}_{i+e_{\nu},\downarrow}(c_{i+e_x\downarrow}+c_{i+e_y\downarrow}),
\end{eqnarray}
where $\rho_{\pm x}=1$, $\rho_{\pm y}=-1$, as given in Eq. (\ref{twoparticle2d}).

\section{Jump operators for the fixed-phase state}\label{appendixb}

In this Appendix, we discuss the general formalism for the construction of jump operators for the fixed-phase state, starting from number-conserving jump operators with known unique dark state with fixe particle number. These jump operators describe dissipative processes for which the total particle number is not exactly conserved, whereas the average particle number approaches the steady state value determined by the parameters of the dissipative process. In the following, we will first discuss the pairing states of spinful fermions, before extending the formalism to spinless fermions.

\subsection{Pairing states with spins}
We only consider separable pairing states, i.e. pairing states whose spin degrees of freedom can be factorised. Then the general number-conserving pairing state for spinful fermions can be written as:
\begin{eqnarray}\label{appbeq1}
|\psi\rangle&=&\left(\sum_i C^{\dag}_{i}A^{\dag}_{i}\right)^N|{\rm vac}\rangle\nonumber\\
&=&\left[\sum_{i}\left(\sum_{\nu}\rho_{\nu}c_{i+\mathbf{e}_{\nu},\uparrow}^{\dag}\right)\left(\sum_{\mu}\lambda_{\mu}c^{\dag}_{i+\mathbf{e}_{\mu},\downarrow}\right)\right]^{N}|{\rm vac}\rangle,
\end{eqnarray}
where $C^{\dag}_i=\sum_{\nu}\rho_{\nu}c^{\dag}_{i+\mathbf{e}_{\nu},\uparrow}$ and $A^{\dag}_i=\sum_{\mu}\lambda_{\mu}c^{\dag}_{i+\mathbf{e}_{\mu},\downarrow}$ are translation invariant. Fourier transform the pairing operators into the momentum space,
\begin{eqnarray}
&&C^{\dag}_{\mathbf{k}}=\sum_i e^{i\mathbf{k}\cdot\mathbf{r}_i}C^{\dag}_i=f_{\mathbf{k}}c^{\dag}_{\mathbf{k},\uparrow},\\
&&A^{\dag}_{\mathbf{k}}=\sum_i e^{i\mathbf{k}\cdot\mathbf{r}_i}A^{\dag}_i=g_{\mathbf{k}}c^{\dag}_{\mathbf{k},\downarrow},
\end{eqnarray}
where $f_{\mathbf{k}}=\sum_{\nu}\rho_{\nu}e^{-i\mathbf{k}\cdot\mathbf{e}_{\nu}}$ and $g_{\mathbf{k}}=\sum_{\mu}\lambda_{\mu}e^{-i\mathbf{k}\cdot\mathbf{e}_{\mu}}$. With these,
the fixed-phase correspondence of Eq. (\ref{appbeq1}) can be written in the form of a coherent state:
\begin{eqnarray}\label{fixphasestate}
|\psi\rangle_c&=&{\cal N}\sum_n\frac{\alpha^n\left(\sum_iC^{\dag}_iA^{\dag}_i\right)^n}{n!}|{\rm vac}\rangle\nonumber\\
&=&{\cal N}\exp\left(\alpha \sum_iC^{\dag}_iA^{\dag}_i\right)|{\rm vac}\rangle\nonumber\\
&=&{\cal N}\prod_{\mathbf{k}}\left(1+\alpha f_{\mathbf{k}}g_{-\mathbf{k}}c^{\dag}_{\mathbf{k},\uparrow}c^{\dag}_{-\mathbf{k},\downarrow}\right)|{\rm vac}\rangle,
\end{eqnarray}
where $\alpha$ is a complex number carrying the phase of the pairing state, ${\cal N}$ is the normalization factor, and $\mathbf{q}$ runs over the first Brillouin zone in the product. Without loss of generality, this coherent state can be re-arranged into the standard BCS-type mean-field wave function:
\begin{equation}\label{appbeqcoh}
|\psi\rangle_c=\prod_{\mathbf{k}}\left(u_{\mathbf{k}}+v_{\mathbf{k}}c^{\dag}_{\mathbf{k},\uparrow}c^{\dag}_{-\mathbf{k},\downarrow}\right)|{\rm vac}\rangle,
\end{equation}
where the coefficients $u_{\mathbf{k}}=\frac{1}{\sqrt{1+|\alpha f_{\mathbf{k}}g_{-\mathbf{k}}|^2}}$, $v_{\mathbf{k}}=\frac{\alpha f_{\mathbf{k}}g_{-\mathbf{k}}}{\sqrt{1+|\alpha f_{\mathbf{k}}g_{-\mathbf{k}}|^2}}$.

Apparently, the coherent state Eq. (\ref{appbeqcoh}) is the vacuum for the Bogoliubov quasiparticle operators:
\begin{eqnarray}
\gamma_{\mathbf{k},\uparrow}&=&u_{\mathbf{k}}c_{\mathbf{k},\uparrow}-v_{\mathbf{k}}c^{\dag}_{-\mathbf{k},\downarrow},\label{g1}\\
\gamma_{\mathbf{k},\downarrow}&=&u_{\mathbf{k}}c_{-\mathbf{k},\downarrow}+v_{\mathbf{k}}c^{\dag}_{\mathbf{k},\uparrow},\label{g2}
\end{eqnarray}
as it is easy to verify the following relations, $\gamma_{\mathbf{k},\uparrow}|\psi\rangle_c=\gamma_{\mathbf{k},\downarrow}|\psi\rangle_c=\gamma_{-\mathbf{k},\uparrow}|\psi\rangle_c=\gamma_{-\mathbf{k},\downarrow}|\psi\rangle_c=0$. Therefore, these Bogoliubov quasiparticle operators are the momentum space jump operators for the fixed-phase state Eq. (\ref{appbeqcoh}). Based on Eqs. (\ref{g1}, \ref{g2}), it is easy to find a more general form of the momentum space jump operators
\begin{eqnarray}
\gamma_{\mathbf{k},\uparrow}&=&\varphi^{+}(\mathbf{k})\left(c_{\mathbf{k},\uparrow}-\frac{v_{\mathbf{k}}}{u_{\mathbf{k}}}c^{\dag}_{-\mathbf{k},\downarrow}\right),\label{gnewm1}\\
\gamma_{\mathbf{k},\downarrow}&=&\varphi^{-}(\mathbf{k})\left(c_{-\mathbf{k},\downarrow}+\frac{v_{\mathbf{k}}}{u_{\mathbf{k}}}c^{\dag}_{\mathbf{k},\uparrow}\right),\label{gnewm2}
\end{eqnarray}
where $\varphi^{\pm}(\mathbf{k})$ are arbitrary functions of $\mathbf{k}$. Fourier transforming Eqs. (\ref{gnewm1}, \ref{gnewm2}) back to the coordinate space, we immediately get the quasi-local jump operators that we look for.

As an illustrating example, let us investigate the simple case with $g_{\mathbf{k}}=1$, $\varphi^{\pm}(\mathbf{k})=1$, which implies the structure of the pairing state should be completely encoded in the spin-up degrees of freedom. The coherent state in this case becomes
\begin{equation}
|\psi\rangle_c={\cal N}\prod_{\mathbf{k}}\left(1+\alpha f_{\mathbf{k}}c^{\dag}_{\mathbf{k},\uparrow}c^{\dag}_{-\mathbf{k},\downarrow}\right)|{\rm vac}\rangle.
\end{equation}
The corresponding momentum space jump operators are:
\begin{eqnarray}
\gamma_{\mathbf{k},\uparrow}&=&c_{\mathbf{k},\uparrow}-\alpha f_{\mathbf{k}}c^{\dag}_{-\mathbf{k},\downarrow},\\
\gamma_{-\mathbf{k},\downarrow}&=&c_{\mathbf{k},\downarrow}+\alpha f_{-\mathbf{k}}c^{\dag}_{-\mathbf{k},\uparrow}.
\end{eqnarray}
We then Fourier transform the momentum space jump operators to the coordinate space,
\begin{eqnarray}
\gamma_{i,\uparrow}&=&c_{i,\uparrow}-\alpha \sum_{\nu}\rho_{\nu}c^{\dag}_{i-\mathbf{e}_{\nu},\downarrow},\\
\gamma_{i,\downarrow}&=&c_{i,\downarrow}+\alpha \sum_{\nu}\rho_{\nu} c^{\dag}_{i+\mathbf{e}_{\nu},\uparrow}. \label{appjump2}
\end{eqnarray}
For the d-wave pairing state, $\rho_{\pm x}=-\rho_{\pm y}=1$, and we recover the fixed-phase jump operators for the d-wave pairing state in Sec. \ref{fixphasesec}. As a consistency check, we demonstrate below that the coherent state in Eq. (\ref{fixphasestate}) is a dark state of the jump operator in Eq. (\ref{appjump2}):
\begin{eqnarray}\label{samplecal}
&&\gamma_{j,\downarrow}\sum_{n}\frac{\alpha^n (\sum_iC^{\dag}_iA^{\dag}_i)^n}{n!}|{\rm vac}\rangle\nonumber\\
&&=\alpha\sum_{n,\nu}\rho_{\nu}c^{\dag}_{j,\uparrow}\frac{\alpha^n (\sum_iC^{\dag}_iA^{\dag}_i)^n}{n!}|{\rm vac}\rangle
-\sum_{n,\nu}\frac{\alpha^n}{n!}
\left[n\rho_{\nu}c^{\dag}_{j+\mathbf{e}_{\nu},\uparrow}(\sum_iC^{\dag}_iA^{\dag}_i)^{n-1}\right]|{\rm vac}\rangle\nonumber\\
&&=\left[\alpha\sum_{\nu}\rho_{\nu}c^{\dag}_{j+\mathbf{e}_{\nu},\uparrow}-\alpha\sum_{\nu}\rho_{\nu}c^{\dag}_{j+\mathbf{e}_{\nu},\uparrow}\right]\sum_{n}\frac{\alpha^n (\sum_iC^{\dag}_iA^{\dag}_i)^n}{n!}|{\rm vac}\rangle\nonumber\\
&&=0.
\end{eqnarray}

Finally, we note that the fixed-phase jump operators presented here are superpositions of a creation operator and an annihilation operator, and hence do not conserve the total particle number. The average particle number on the other hand, is driven to the final steady state value during the dissipative process. The average particle number in the final steady state is given as:
\begin{equation}
N=\sum_{\mathbf{k},\sigma}\langle c^{\dag}_{\mathbf{k},\sigma}c_{\mathbf{k},\sigma}\rangle=\sum_{\mathbf{k}}\frac{2|\alpha f_{\mathbf{k}}|^2}{1+|\alpha f_{\mathbf{k}}|^2}.
\end{equation}

\subsection{Spinless pairing states}
One can easily extend the formalism above to the pairing states of spinless fermions, with modifications to the pairing parameters due to the triplet pairing symmetry. The pairing state with fixed phase can be written as
\begin{eqnarray}
|\psi\rangle_p&=&{\cal N}\exp\left[\alpha \sum_i\left(\sum_{\nu}\rho_{\nu}c^{\dag}_{i+\mathbf{e}_{\nu}}\right)\left(\sum_{\mu}\lambda_{\mu}c^{\dag}_{i+\mathbf{e}_{\mu}}\right)\right]|{\rm vac}\rangle\nonumber\\
&=&\prod_{\mathbf{k}}\displaystyle'\left(u_{\mathbf{k}}+v_{\mathbf{k}}c^{\dag}_{\mathbf{k}}c^{\dag}_{-\mathbf{k}}\right)|{\rm vac}\rangle,
\end{eqnarray}
where $\prod'$ indicates that $\mathbf{k}$ only runs over half of the Brillouin zone, e.g. $k_x>0$. The coefficients here are similar to the spinful case, with $u_{\mathbf{k}}=\frac{1}{\sqrt{1+|2\alpha f_{\mathbf{k}}g_{-\mathbf{k}}|^2}}$, $v_{\mathbf{k}}=\frac{2\alpha f_{\mathbf{k}}g_{-\mathbf{k}}}{\sqrt{1+|2\alpha f_{\mathbf{k}}g_{-\mathbf{k}}|^2}}$, $f_{\mathbf{k}}=\sum_{\nu}\rho_{\nu}e^{-i\mathbf{k}\cdot\mathbf{e}_{\nu}}$ and $g_{\mathbf{k}}=\sum_{\mu}\lambda_{\mu}e^{-i\mathbf{k}\cdot\mathbf{e}_{\mu}}$.
As a simple example, we consider the case with $g_{-\mathbf{k}}=1$. The coherent state then becomes
\begin{equation}
|\psi\rangle_p={\cal N}\prod_{\mathbf{k}}\displaystyle'\left(1+2\alpha f_{\mathbf{k}}c^{\dag}_{\mathbf{k}}c^{\dag}_{-\mathbf{k}}\right)|{\rm vac}\rangle,
\end{equation}
where due to the triplet pairing symmetry, $f_{-\mathbf{k}}=f^{\ast}_{\mathbf{k}}=-f_{\mathbf{k}}$.

Following the previous approach, the general form of the momentum space jump operators are
\begin{equation}
\gamma_{\mathbf{k}}=\varphi(\mathbf{k})(c_{\mathbf{k}}-2\alpha f_{\mathbf{k}} c^{\dag}_{-\mathbf{k}}),\label{pg1}
\end{equation}
where $\varphi(\mathbf{k})$ is an arbitrary function of $\mathbf{k}$. For simplicity, we take $\varphi(\mathbf{k})=1$, and the Fourier transform of Eq. (\ref{pg1}) gives the quasi-local jump operators:
\begin{equation}
\gamma_{i}=c_{i}+2\alpha \sum_{\nu}\rho_{\nu}c^{\dag}_{i+\mathbf{e}_{\nu}}.
\end{equation}
Similar to Eq. (\ref{samplecal}), it is straightforward to check that $\gamma_i|\psi\rangle_p=0$.

Finally, the average particle number is given by
\begin{equation}
N=\sum_{\mathbf{k}}\langle c^{\dag}_{\mathbf{k}}c_{\mathbf{k}}\rangle=\sum_{\mathbf{k}}\frac{|2\alpha f_{\mathbf{k}}|^2}{1+|2\alpha f_{\mathbf{k}}|^2}£¬
\end{equation}
where the summation runs over the entire first Brillouin zone. With $\rho_{x}=-\rho_{-x}=-\mathrm i\rho_{y}= \mathrm i \rho_{-y} =1$, we recover the results for p-wave pairing states in Sec. \ref{fixphasesec}.

\section{Derivation of the effective decay rate}\label{appendixc}

In this Appendix, we derive the effective decay rate from $^3$P$_0$ to $^1$S$_0$ manifold via cavity mode during the implementation scheme illustrated in Sec. VI. The level scheme and various parameters are shown in Fig. (\ref{Figlevel}). We basically need to adiabatically eliminate the intermediate states in the $^1$P$_1$ manifolds as well as the cavity photon mode sequentially.

For the clarity of discussion, we denote $|a\rangle=|^3{\rm P}_0,0\rangle$, $|b\rangle=|^1P_1,0\rangle$,$|c\rangle=|^1{\rm S}_0,1\rangle$,
$|d\rangle=|1^S_0,0\rangle$, where the second index indicates the number of photons in the cavity mode, and we have neglected the indices for the hyperfine spins for simplicity. The master equation for the density matrix is then
\begin{equation}
\dot{\rho}=-i[H,\rho]-\frac{\Gamma}{2}(a^{\dag}a\rho+\rho a^{\dag}a)+\Gamma a\rho a^{\dag},
\end{equation}
where $a$($a^{\dag}$) is annihilation (creation) operator for the cavity photon mode. The Hamiltonian under the rotating wave approximation and appropriate rotating frame reads:
\begin{equation}
H=\Delta a^{\dag}_b a_b+ (\frac{\Omega}{2}a^{\dag}_a a_b+h.c.)+(g a^{\dag}_{b}a_c+h.c.),
\end{equation}
where $a_i$(i=a,b,c,d) is the annihilation operator for the corresponding level, $\Omega$ is the effective Rabi-frequency between $|a\rangle$ and $|b\rangle$, and $g$ is the coupling rate between the cavity and the atom.

The equations of motion become:
\begin{eqnarray}
\dot{\rho}_{aa}&=&-i\frac{\Omega}{2}(\rho_{ba}-\rho_{ab}) \\
\dot{\rho}_{bb}&=&-i\frac{\Omega}{2}(\rho_{ab}-\rho_{ba})-ig(\rho_{cb}-\rho_{bc})\\
\dot{\rho}_{cc}&=&-ig(\rho_{bc}-\rho_{cb})-\Gamma\rho_{cc}\\
\dot{\rho}_{dd}&=&\Gamma\rho_{cc}\\
\dot{\rho}_{ab}&=&i\Delta\rho_{ab}+ig\rho_{ac}+i\frac{\Omega}{2}(\rho_{aa}-\rho_{bb})\\
\dot{\rho}_{ac}&=&-i\frac{\Omega}{2}\rho_{bc}+ig\rho_{ab}-\frac{\Gamma}{2}\rho_{ac}\\
\dot{\rho}_{bc}&=&-i\Delta\rho_{bc}-i\frac{\Omega}{2}\rho_{ac}-ig(\rho_{cc}-\rho_{bb})-\frac{\Gamma}{2}\rho_{bc}\\
\dot{\rho}_{cd}&=&-ig\rho_{bd}-\frac{\Gamma}{2}\rho_{cd}\\
\dot{\rho}_{bd}&=&-i\Delta\rho_{bd}-ig\rho_{cd}-i\frac{\Omega}{2}\rho_{ad}\\
\dot{\rho}_{ad}&=&-i\frac{\Omega}{2}\rho_{bd}
\end{eqnarray}
where $\rho_{ij}=\langle i|\rho|j\rangle$. Physically, the detuning from $^1$P$_1$ manifold should be much larger than the its bandwidth to avoid large spontaneous emission, therefore $\Delta\gg 28$MHz becomes the largest energy scale in the equations. We may then adiabatically eliminate $^1$P$_1$ manifold first, which amounts to setting $\dot{\rho}_{bd}=\dot{\rho}_{bc}=\dot{\rho}_{ba}=0$. The resulting equations of motion become:
\begin{eqnarray}
\dot{\rho}_{aa}&=&-i\frac{\Omega g}{2\Delta}(\rho_{ac}-\rho_{ca})\\
\dot{\rho}_{cc}&=&i\frac{\Omega g}{2\Delta}(\rho_{ac}-\rho_{ca})-\Gamma\rho_{cc}\\
\dot{\rho}_{dd}&=&\Gamma\rho_{cc}\\
\dot{\rho}_{ac}&=&i\frac{\Omega g}{2\Delta}(\rho_{cc}-\rho_{aa})+i(\frac{\Omega^2}{4\Delta}-\frac{g^2}{\Delta})\rho_{ac}-\frac{\Gamma}{2}\rho_{ac}\\
\dot{\rho}_{ad}&=&i\frac{\Omega g}{2\Delta}\rho_{cd}+i\frac{\Omega^2}{4\Delta}\rho_{ad}\\
\dot{\rho}_{cd}&=&i\frac{g^2}{\Delta}\rho_{cd}+i\frac{\Omega g}{2\Delta}\rho_{ad}-\frac{\Gamma}{2}\rho_{cd},
\end{eqnarray}
where we have assumed $\Delta\gg \Gamma,\Omega$ and neglected terms on the order of $(\frac{g}{\Delta})^2,(\frac{\Omega}{\Delta})^2$.

We may then adiabatically eliminate the cavity photon mode by $\dot{\rho}_{ac}=\dot{\rho}_{dc}=0$. This way, we arrive at the final equations of motion
\begin{eqnarray}
\dot{\rho}_{ad}&=&-\frac{\Gamma_{\mathrm{eff}}}{2}\rho_{ad}+i\frac{\Omega^2}{4\Delta}\rho_{ad}\\
\dot{\rho}_{aa}&=&-\Gamma_{\mathrm{eff}}\rho_{aa}\\
\dot{\rho}_{dd}&=&\Gamma_{\mathrm{eff}}\rho_{dd}
\end{eqnarray}
where the effective decay rate $\Gamma_{\mathrm{eff}}=\frac{\Omega^2g^2}{\Delta^2\Gamma}$, and we have assumed $\Gamma\gg \max( \frac{\Omega^2}{\Delta},\frac{g^2}{\Delta},\frac{\Omega g}{\Delta})$.  For typical experimental parameters:  $\Delta\sim100$MHz, $\Omega\sim10$MHz, $\kappa\sim10$MHz, $g\sim3$MHz, we obtain an effective decay rate $\Gamma_{\mathrm{eff}}\sim9$kHz.

\end{document}